\documentclass[aps,prd,twocolumn,nofootinbib,showpacs]{revtex4}

\usepackage{rotating}
\usepackage{graphicx}
\usepackage{epsfig} 
\usepackage{graphicx}
\usepackage{bm}

\newcommand{\zjets }{$Z$+jets}
\newcommand{\wjets }{$W$+jets}
\newcommand{\gammajets }{$\gamma$+jets}
\newcommand{\ttbarjets }{$t \bar{t}$}
\newcommand{\pt }{$|\vec{p}_T|$}
\newcommand{\ptlead }{$|\vec{p}_T^{\rm \;lead}|$}
\newcommand{\met}{$E_T^{\rm miss}$}

\begin{document}

\title{ Background Modeling in New Physics Searches Using Forward Events at LHC }

\author{Victor Pavlunin}
\author{David Stuart}
\affiliation{Department of Physics, University of California, Santa Barbara, CA, USA, 93106-9530 }

\date{\today}

\begin{abstract}

     We present a method to measure dominant Standard Model~(SM) backgrounds 
     using data containing high rapidity objects in $pp$ collisions at 
     the Large Hadron Collider~(LHC). 
     The method is developed for analyses of early LHC data when 
     robustness against imperfections of background modeling  and detector 
     simulation can be a key to the discovery of new physics at LHC.

\end{abstract}

\maketitle

\vspace{10mm}

\section{Introduction}

The LHC will soon start operating in an unexplored energy regime at 
$\sqrt{s} \sim 14$~TeV, about seven times higher than that achieved 
at the Tevatron. At that center-of-mass energy, a large number of new particles 
could be produced even in a data sample of modest integrated luminosity.
The challenge is to distinguish events with new particles from those, 
many orders of magnitude more copious, attributed to the SM,
and to do so using tools and methods appropriate for early data.
The challenge is magnified by the fact that signatures 
of the physics beyond the SM realized in nature are not known.

Heavy new particles are produced,  approximately at threshold, 
via interactions of energetic partons.
Their decay products tend to be distributed  
uniformly over solid angle, which corresponds to 
a narrow central rapidity region~\cite{rapidity}.  
SM particles are light on the mass scale of $14$~TeV 
and tend to be produced in interactions of soft, often very asymmetric 
in energy, partons. They receive a significant boost along the beam line,
which makes them distributed over a wide rapidity range.

In this paper, we present a new method to measure dominant SM 
backgrounds in searches for heavy new particles. 
It uses data containing high rapidity objects to predict SM yields at 
small rapidity. We apply this to the SM processes:  
\zjets, \wjets, \gammajets, QCD jets and \ttbarjets, that are 
the largest background sources in many new physics searches.
We also discuss the usage of a ratio constructed from event yields 
in central and forward rapidity regions as a generic search variable.

The method is presented in the context of a new physics search 
involving leptons, photons, jets and missing transverse  energy.
In the absence of a single most compelling model of new physics, 
the search is developed in a model independent way. 
The only assumption we make is that new particles 
are heavy and they decay to SM particles via a multi-stage 
cascade producing a large number of jets, so that the number 
of jets is a main search variable. 
A key feature of our method  is that systematic uncertainties associated 
with incomplete knowledge of the SM production rates and detector artifacts
cancel to first order. The emphasis throughout 
is on robustness against imperfections of background modeling  
required for new physics searches in early LHC data.

\section{Method Overview}

\label{method}

We consider final states involving many jets, 4 or more.
The SM $V+$jets production rates, 
where for brevity $V$ stands for a $Z$, $W$, $\gamma$ or a jet~\cite{ttbarlater}, 
fall steeply  as the number of jets grows, but they are difficult 
to predict from first principles. 
Monte Carlo~(MC) techniques are unreliable in predicting backgrounds with a large number of jets. 
Theory calculations~\cite{theory_predictions} do not exist at sufficiently high order.
The structure functions have significant uncertainties
for partons carrying a small fraction, $x$, of the proton 
momentum that is relevant for LHC~\cite{pdf_uncert}.  
Large uncertainties in the 
calibration of the experimental apparatus are expected in early 
data taking. For these reasons, instead of relying on MC simulation 
of the detector response to SM processes, we use control regions in data to determine 
dominant SM backgrounds. 
We identify control samples in kinematic regimes 
where the SM dominates and extrapolate backgrounds measured there
into the signal region where new physics may contribute. 
In $V+$jets, the SM dominates when the transverse momentum, \pt, of $V$ 
or the number of jets, $N_J$, is small.
These control regions have been used previously for data-based background 
determination~\cite{other-methods}. 
We use, in addition, control samples with high rapidity objects 
that are background dominated even when \pt~or $N_J$ is large.
Jet rapidity has been successfully used previously in di-jet 
resonance searches at the Tevatron~\cite{dijetratios}.

\begin{figure*}[htbp]
\begin{center}
\begin{minipage}{5.4in}
\epsfig{file=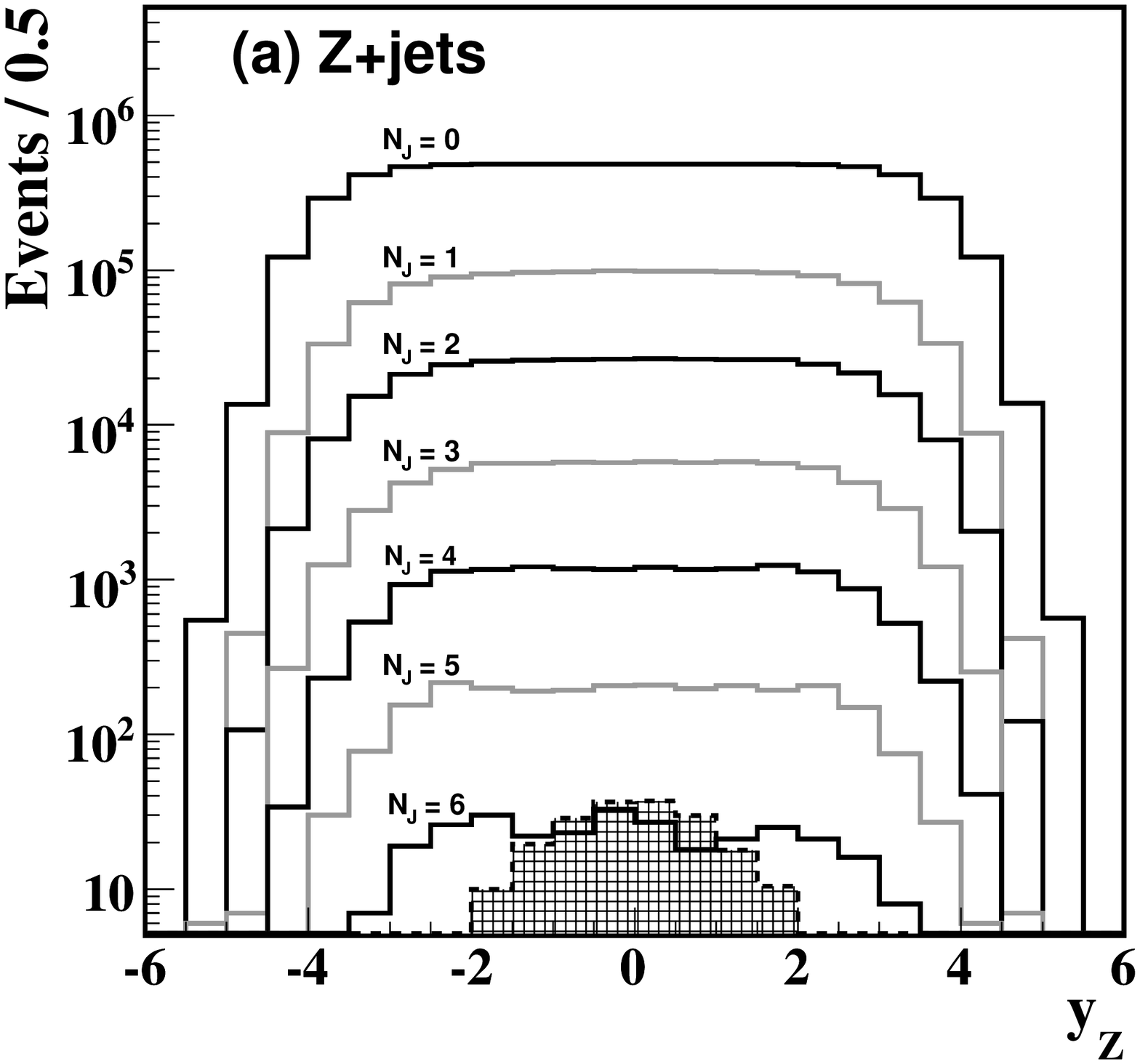,width=2.6in}
\epsfig{file=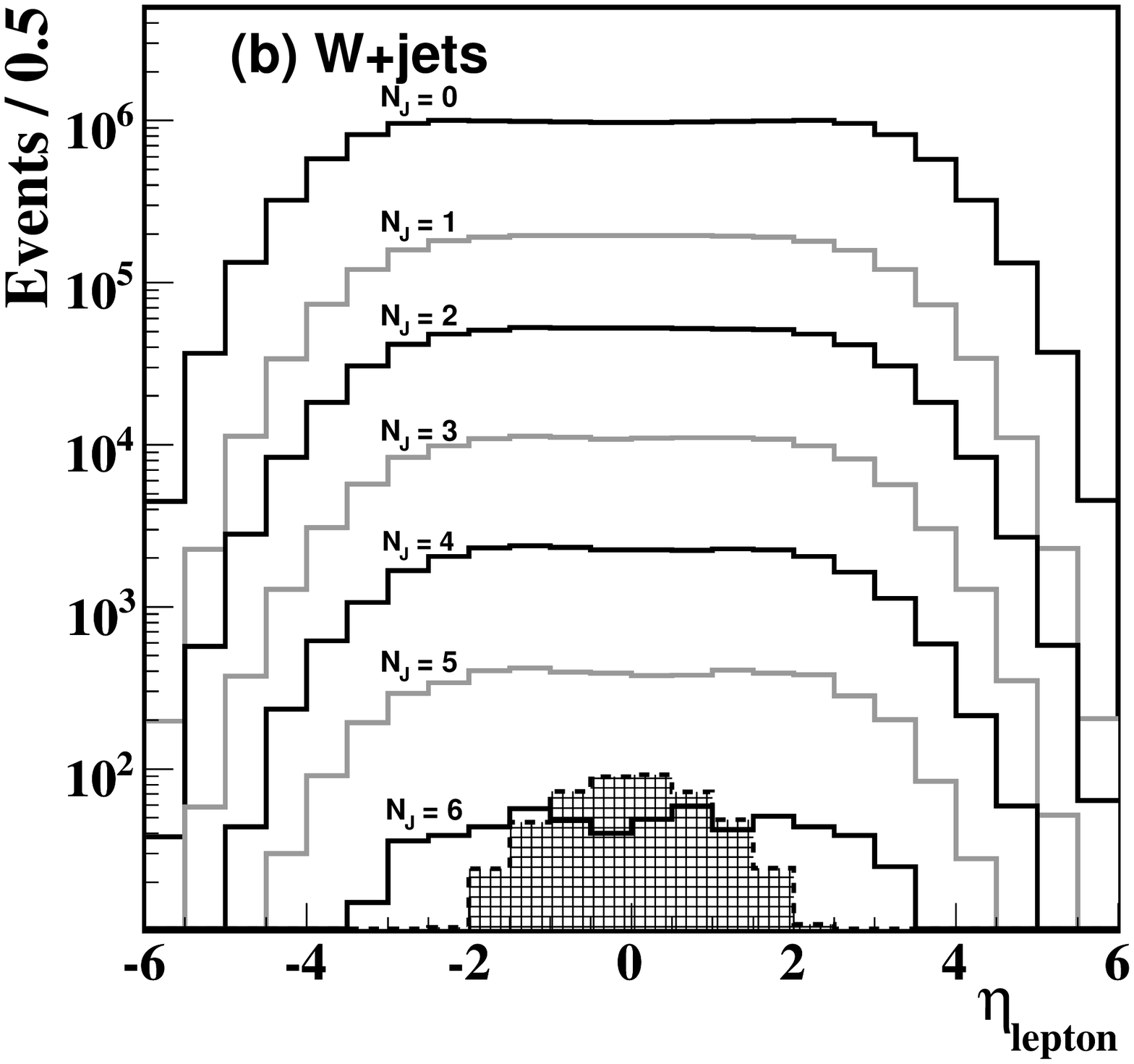,width=2.6in}
\vspace{-2mm}
\end{minipage}
\begin{minipage}{5.4in}
\epsfig{file=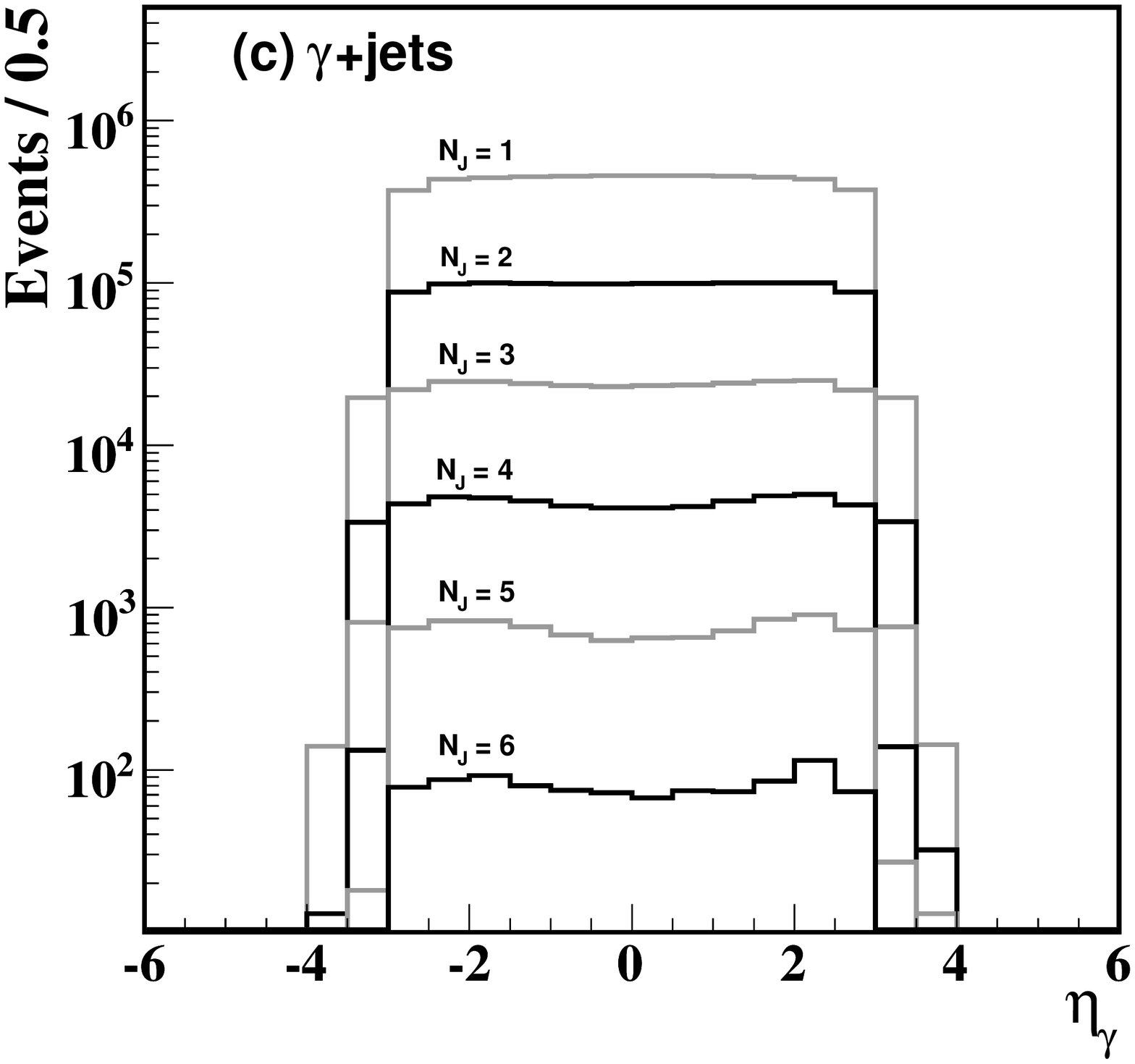,width=2.6in}
\epsfig{file=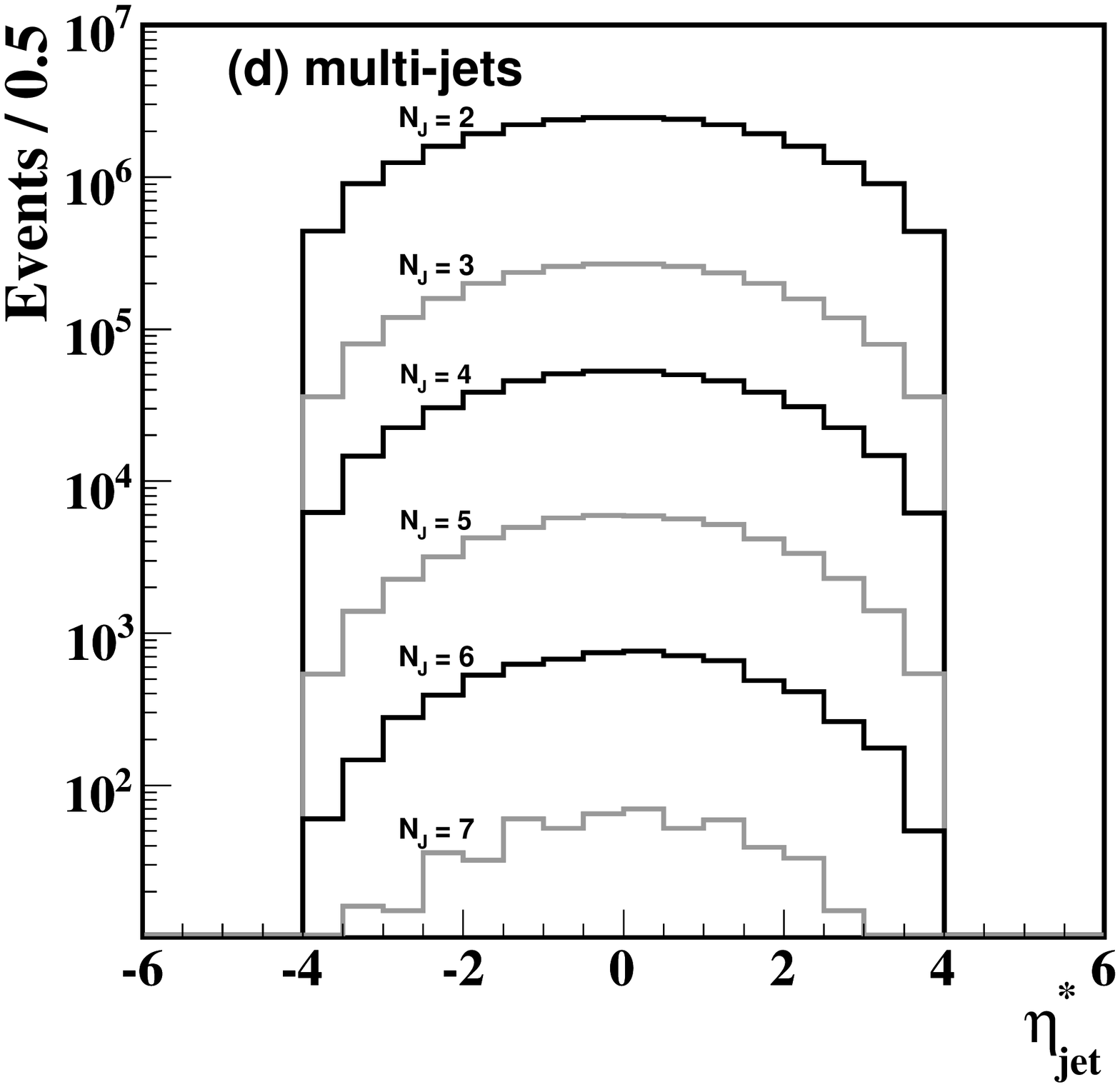,width=2.6in}
\end{minipage}
\end{center}
\caption{ 
      Rapidity of $Z$-bosons from SM \zjets~(a), 
      pseudo-rapidity of charged leptons from SM~\wjets~(b),
      pseudo-rapidity of $\gamma$ for SM \gammajets~(c), and
      pseudo-rapidity of the highest \pt~jet in SM QCD multi-jet events~(d).
      Generator level requirements of $|\eta_\gamma| < 3.0$ and $|\eta_{\rm jet}| < 4.0$ are imposed 
      in plots~(c) and~(d).
      Shapes of rapidity distributions from LM4 and LM6 mSUGRA benchmark points~\cite{cms-msugra}
      are shown by black hatched histograms in the \zjets~and \wjets~cases, respectively. 
      }
\label{figure-1}
\end{figure*}

Figure~\ref{figure-1} shows the (pseudo-)rapidity distributions for
\zjets~(a), \wjets~(b), \gammajets~(c), and multi-jets~(d).
In the \zjets~channel, we use the rapidity of the $Z$ boson, $y_Z$, 
as a key discriminating rapidity variable.
The $W$ boson rapidity cannot be unambiguously determined due to the undetected neutrino. 
We instead use the lepton pseudo-rapidity~\cite{rapidity}, $\eta_{\rm lepton}$, for \wjets. 
The pseudo-rapidities of the photon, $\eta_{\gamma}$, and the highest 
\pt~jet, $\eta_{\rm jet}^*$, are used for \gammajets~ and multi-jets, respectively.
As seen in Figure~\ref{figure-1}, the (pseudo-)rapidity 
distributions for decays of new massive particles are central,
while that for the SM processes are approximately 
uniform in a wide rapidity range. 
Furthermore, the rapidity distributions vary slowly as 
the number of jets increases. 

The object providing the discriminating rapidity variable is called a tag~\cite{other-tags}.
We use events with forward tags to determine backgrounds for events 
with central tags, using an algorithm described in section~\ref{algorithm}.

In this paper, for brevity, we discuss searches at high $N_J$, since 
$N_J$  is a particularly simple and robust variable.
Other distributions considered in our search include:
the highest jet \pt~(\ptlead)~and the $J_{T}\equiv \sum |\vec{p}_{T}^{\rm \;jet}|$ spectra in each $N_J$ bin;
and $N^*_J$ distributions, 
which are closely related to $N_J$ but obtained as a sum of weights 
of either \ptlead~or  $J_T$ in each $N_J$ bin.
The $N^*_J$ distributions have higher discriminating power compared to 
the $N_J$ distributions since new particles are expected to be heavy. 
However, reliance on the \ptlead~or $J_T$ spectra is more  susceptible 
to uncertainties in the jet energy scale.

\section{Experimental Aspects}

\label{experiment} 

The ATLAS and CMS experiments use multi-purpose detectors 
that are in the final stages of construction at the European Organization for Nuclear Research~(CERN). 
Detailed descriptions of the detectors can be found in Ref.~\cite{detectors}. 
Of primary importance for our studies are the detectors' rapidity coverages and kinematic thresholds.
The detectors are capable of efficiently reconstructing electrons and muons
with low fake rates for lepton \pt$\;> 20$~GeV within $|\eta| < 2.5$. 
Photons and jets are reconstructed in the $|\eta| < 2.5$ and $|\eta| < 3.0$ range, respectively. 
Missing transverse  energy, \met, is calculated using $E_T$~measurements of all reconstructed 
objects in each event. Mis-measured or mis-reconstructed objects, calorimeter 
noise, malfunctioning detector subsystems and channels, and background unrelated to 
$pp$ collisions constitute sources of unphysical \met~that may complicate 
the usage of \met~in early searches. 
Accordingly, we perform studies with and without a requirement on \met~ in 
the event selection.

To study the effectiveness of the method, we have produced mock data samples for 
the following SM processes: \zjets~(5.0~fb$^{-1}$, up to 5 partons, $Z \rightarrow l^+ l^-$), 
\wjets~(1.0~fb$^{-1}$, up to 5 partons, $W \rightarrow l \nu_l$), 
\ttbarjets~(1.0~fb$^{-1}$, up to 4 partons, $t\bar{t} \rightarrow l \nu_l bbjj$ and 
$t\bar{t} \rightarrow l \nu_l \tau \nu_\tau  bb$),  
\gammajets~(400.0~pb$^{-1}$, up to 5 partons) and QCD jets 
(1.0~pb$^{-1}$, up to 5 partons), where $l$ is $\mu$ or $e$. 
The integrated luminosity indicated in parentheses for each channel specifies 
the sample size used in our studies, except where specified otherwise. 
These samples were generated with ALPGEN~\cite{alpgen}, 
and PYTHIA~\cite{pythia} was used for parton showering, hadronization, 
simulation of the underlying event and jet reconstruction.
To model features of a new physics signal in search distributions, 
we produced mock signal data samples for Minimal Supergravity (mSUGRA) 
benchmark points LM4 and LM6~\cite{susy,msugra,cms-msugra} using PYTHIA.

Kinematic selection criteria are applied as follows.
Electrons and muons are required to have \pt~of at least 20~GeV in the $|\eta| < 2.5$ range. 
Photons are reconstructed above the \pt~
threshold of 30~GeV in the $|\eta| < 2.5$ range. 
Jets are reconstructed using the PYCELL algorithm~\cite{pythia} and 
required to be within $|\eta| < 3.0$ for 
\pt~thresholds varying between $30$ and $100$~GeV.
Low thresholds are used for background studies, while
higher thresholds are used to study signal dominated regions.

\begin{figure}[h]
\epsfig{file=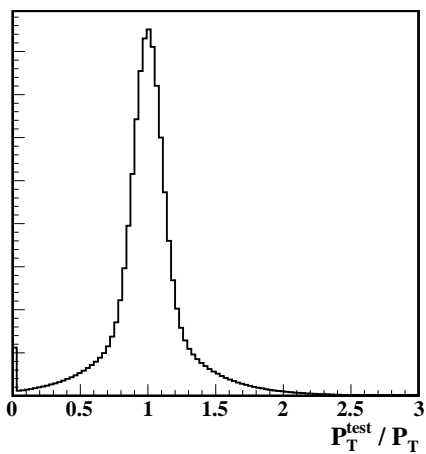,width=2.5in}
\caption{ 
A hypothetical probability density function used for jet energies
in modeling the effect of artificial \met. }
\label{figure-2}
\end{figure}

Detector response is not directly simulated, 
although an assumed reconstruction efficiency of 50\% is applied in each channel.
The \met~ vector is approximated by a vector opposite to the sum of $\vec{p}_T$ measurements 
of charged leptons, photons, and jets. 
Using the \gammajets~ sample, we find that the jet energy resolution function
in our mock data samples is approximately Gaussian with $\sigma$ varying from 
about 15\% at 30 GeV to about 8\% at 100 GeV.
To simulate effects of \met~mis-modeling due to jet energy fluctuations with 
non-Gaussian tails and incomplete hermeticity of the detectors, 
we perform robustness tests where jet energies are varied according to
the hypothetical probability density function shown in Figure~\ref{figure-2},
and jets are removed in selected regions, as described in section~\ref{met}. 

These selection criteria and sample sizes are chosen generally and are not optimized to 
any new physics model. The new physics reference models listed above are used only 
for illustration. Our goal in this paper is to demonstrate the scope of the method
and its performance rather than to attain high sensitivity to a specific model for 
a specific final state or quantify that sensitivity.

\section{ Algorithm }

\label{algorithm}

To describe and illustrate the algorithm and tests of its robustness, 
in the next several sections 
we center the discussion on the \zjets~channel. 
The discussion applies to all four $V+$jets channels, however, 
and differences among these channels are pointed out where significant.

\begin{figure}[h]
\begin{center}
\epsfig{file=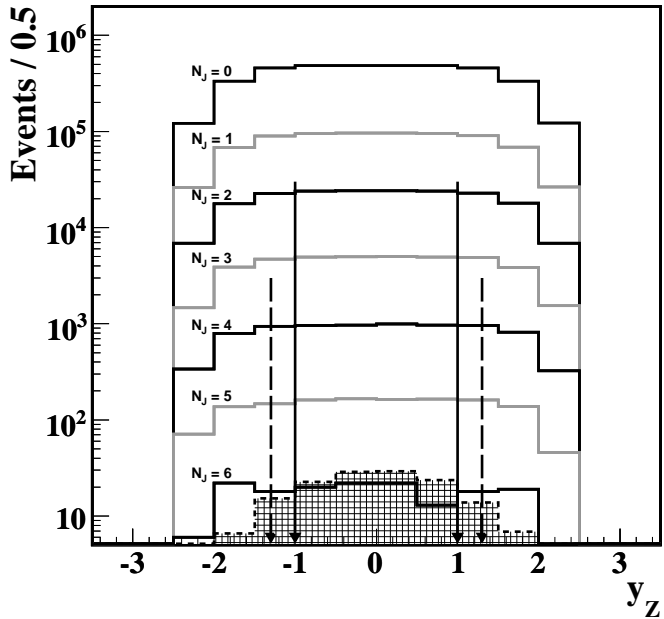,width=3.5in}
\end{center}
\caption{ Rapidity of $Z$-bosons from the SM \zjets~ production in the 
          fiducial coverage of LHC detectors. 
          The central signal region is indicated by solid arrows. The 
          background dominated region is at $|y_Z|$ larger than that
          indicated by dashed arrows. The $Z$ rapidity shape from a LM4
          mSUGRA benchmark is shown by the black hatched histogram. }
\label{figure-3}
\end{figure}

The rapidity range for reconstructed $Z$ bosons passing realistic 
event selection criteria is reduced~(Figure~\ref{figure-3}). 
We define forward events as those with a $Z$ boson having $|y_Z| > 1.3$,
and we call the detector region with $|\eta| > 1.3$ the forward region. 
Central events are defined as those with a $Z$ boson at  $|y_Z| < 1.0$, 
and the central region of the detector as that having $|\eta| < 1.0$. 
(This definition of central and forward categories is arbitrary and could be modified 
without significant effect.) 

Small $N_J$ bins are SM dominated for both central and forward events, 
and we use them to predict the SM contribution to the central, high $N_J$ bins
where signal would appear.
This is done by measuring a ratio, denoted as $R_{N_J}$, 
of the central yield~($Y^{\rm Central}_{N_J}$) to the sum of 
forward~($Y^{\rm Forward}_{N_J}$) and central yields in each $N_J$ bin:
$R_{N_J} \equiv Y^{\rm Central}_{N_J}/(Y^{\rm Forward}_{N_J}+Y^{\rm Central}_{N_J})$.
A linear fit to $R_{N_J}$ is made in the low $N_J$ bins and extrapolated into the high $N_J$ region. 
The extrapolated ratios and the yields of forward events in high $N_J$ bins are combined to obtain 
a background prediction in the central, high $N_J$ signal region.

\begin{figure*}[htb]
\begin{center}
\begin{minipage}{7.1in}
\epsfig{file=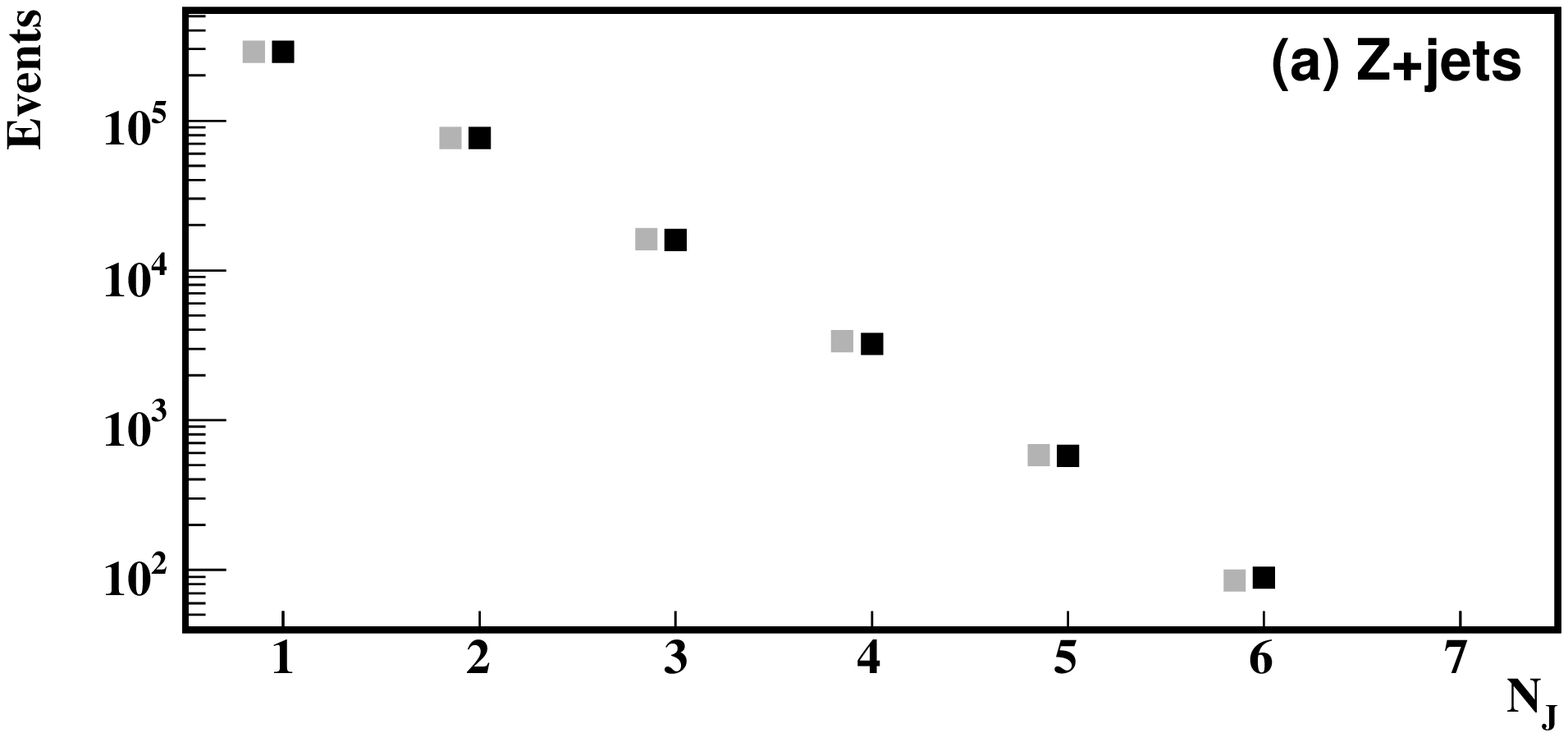,width=3.5in}
\epsfig{file=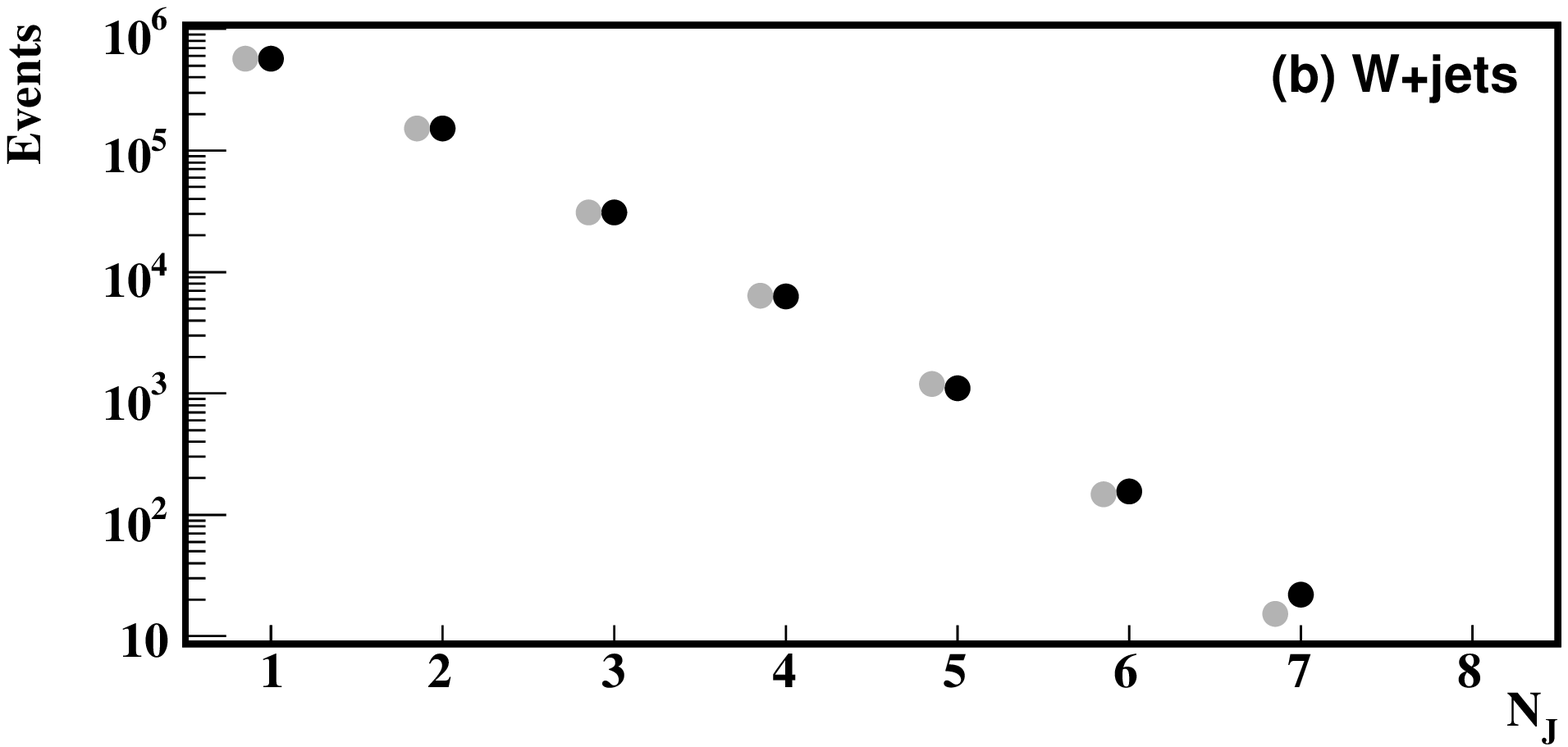,width=3.5in}
\end{minipage}
\begin{minipage}{7.1in}
\epsfig{file=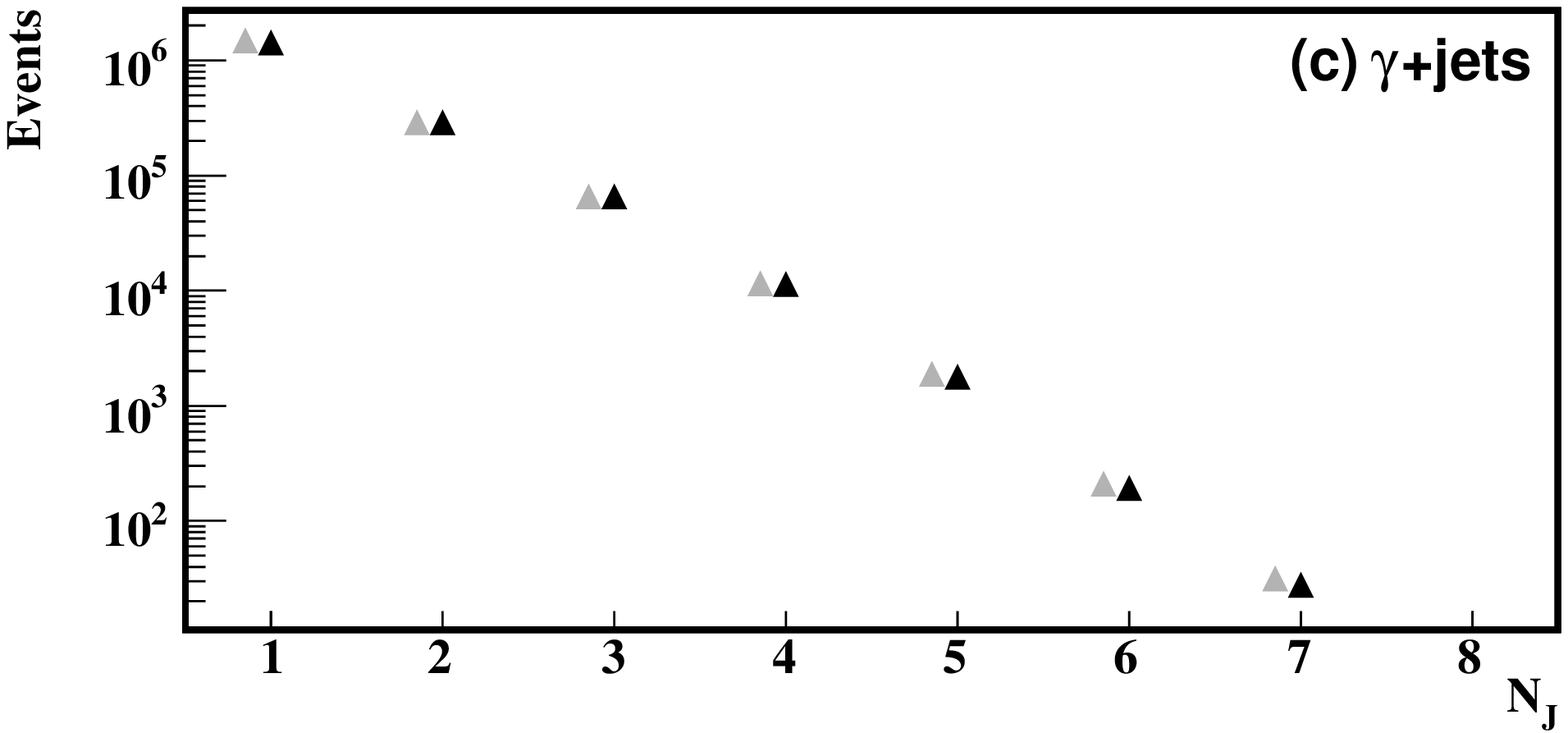,width=3.5in}
\epsfig{file=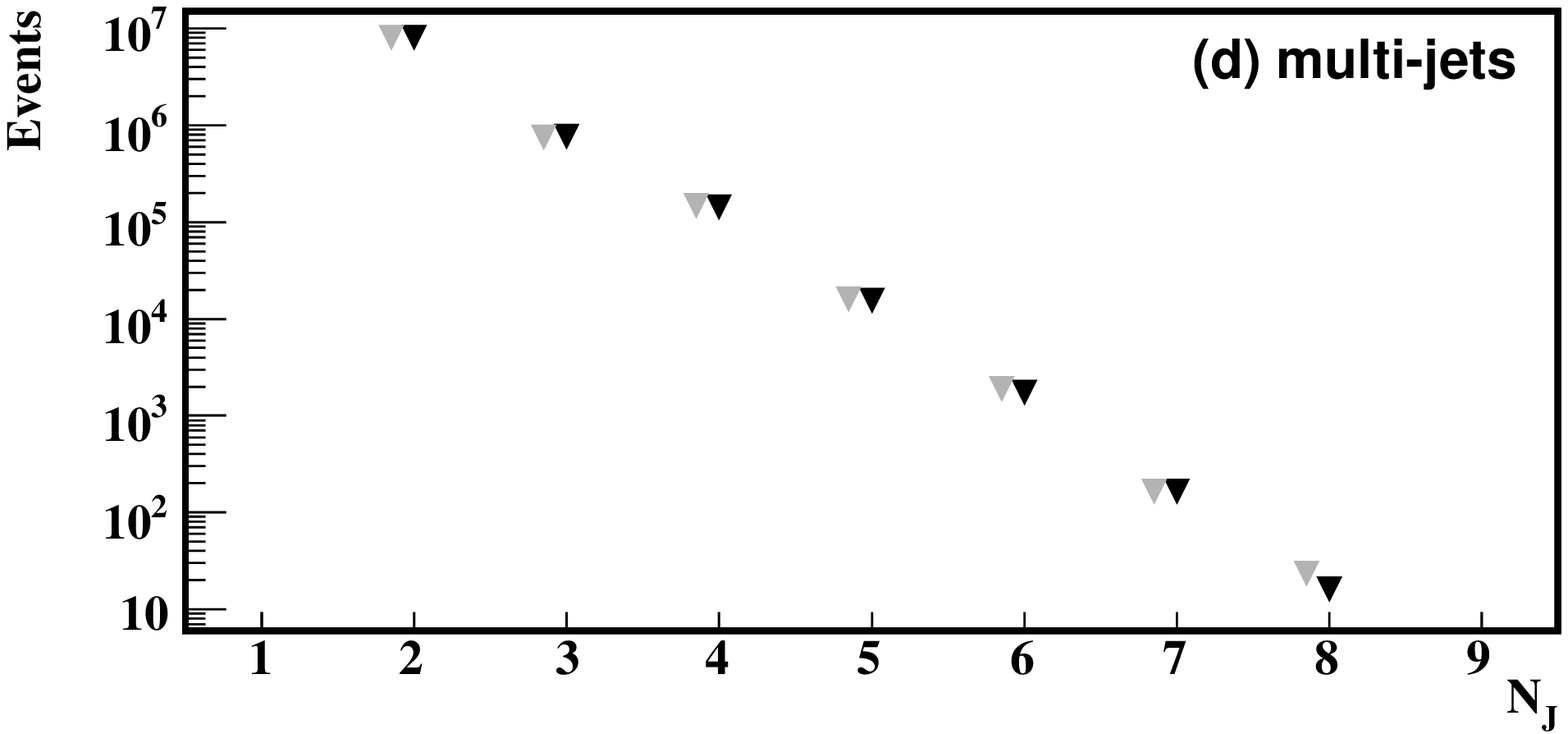,width=3.5in}
\end{minipage}
\begin{minipage}{7.1in}
\epsfig{file=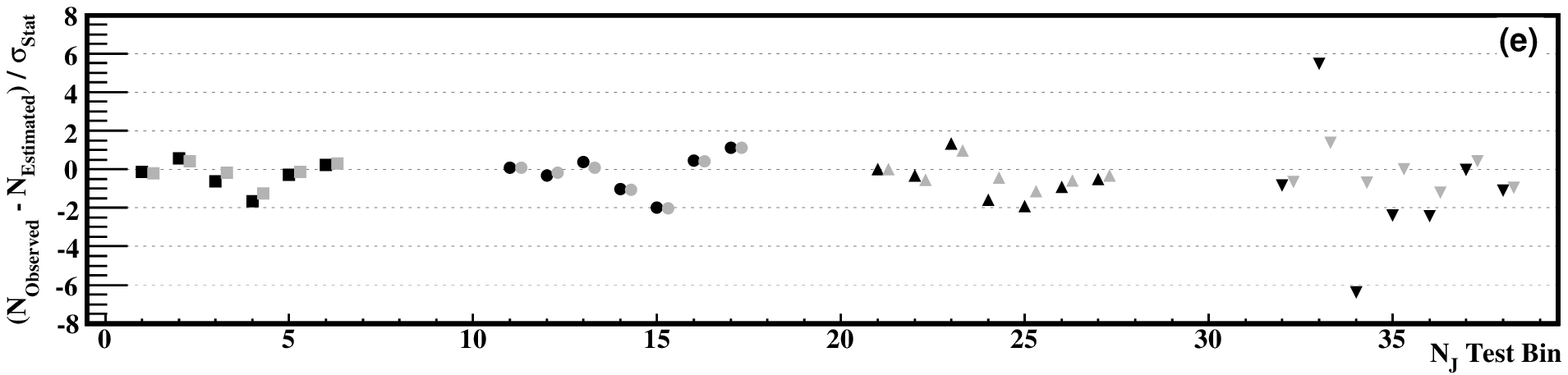,width=7.1in}
\end{minipage}
\end{center}
\caption{ The $N_J$ distributions for \zjets~(a), \wjets~(b),
     	  \gammajets~(c) and pure multi-jets~(d). The backgrounds in the 
       	central regions are shown in black markers, its estimate is in shaded markers of the same
       	shape displaced horizontally for visibility. 
       	Bottom plot: pull distributions for \zjets~(black squares), 
       	\wjets~(black circles), \gammajets~(black triangle-up) and pure multi-jets~(black triangle-down). 
        Here, $N_J$ is offset by 10 between samples for visibility, $i.e.$, $N_J =$ Test Bin mod 10.
        Shaded markers in the bottom plot show how the pulls change after an addition of a 
          1.0\% relative systematic uncertainty in each $N_J$ bin. }
\label{figure-4}
\end{figure*}

The accuracy of this background prediction can be tested in mock data samples by 
comparing it to the yield in the central region at high $N_J$.
This estimated-to-observed comparison is shown as a function of $N_J$ in Figure~\ref{figure-4} 
for \zjets, \wjets, \gammajets~and pure QCD jets.
The prediction is made using fits in $1 \le N_J \le 3$ for \zjets~and \wjets.
For \gammajets~and multi-jets, $2 \le N_J \le 4$ is used.
The observed central yield at high $N_J$ is well matched to the prediction in all cases.
Pull distributions, defined as $(N_{\rm Observed} - N_{\rm Estimated})/\sigma_{\rm Stat}$, 
where $N_{\rm Observed}$ is the observed number of central events,
$N_{\rm Estimated}$ is the number of central events estimated using the algorithm 
and $\sigma_{\rm Stat}$ is the total statistical uncertainty, are in the bottom 
plot of the same Figure in black markers of the appropriate shape for each channel.
Shaded markers in the bottom plot show how the pulls change with 
the addition of a 1\% relative systematic uncertainty in each $N_J$ bin. 
With at most a small systematic uncertainty, the algorithm estimates the background in
the central region accurately.

\begin{figure}
\begin{center}
\vspace{-3mm}
\epsfig{file=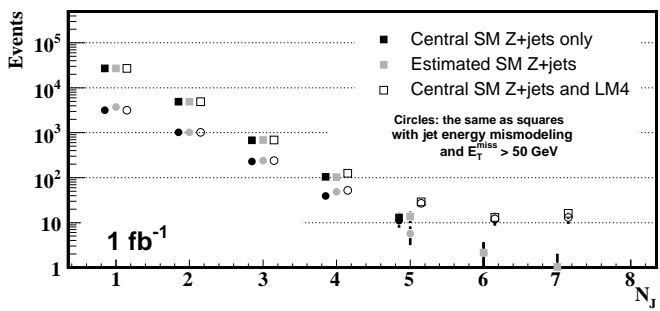,width=3.5in}
\end{center}
\caption{$N_J$~ distributions for \zjets~(black markers) and a mixture of \zjets~ and events 
from LM4 mSUGRA benchmark (shaded markers: estimated central SM background, open markers: all central events).
This comparison is made with a $50$ GeV jet energy threshold and a sample size corresponding to $1~{\rm fb}^{-1}$.
The effect of a \met~$>$~50~GeV requirement for a sample with the jet energy 
mis-modeling discussed in section~\ref{met} is shown by the circles.} 
\label{figure-5}
\end{figure}

The results in Figure~\ref{figure-4} are obtained with a jet threshold of $30$ GeV.
A higher threshold would likely improve signal sensitivity,
but it could also affect the algorithm's performance.
As the jet threshold changes, the $R_{N_J}$ values may change,
but the low $N_J$ fit should properly account for any difference.
We search for the presence of biases by varying the jet threshold between 
30 and 100 GeV and repeating the tests in Figure~\ref{figure-4}~(e) 
for the \zjets~and \wjets~ channels. No evidence of a bias is found.

The performance of the algorithm when signal is present 
is illustrated in Figure~\ref{figure-5}, where we compare the central 
yields and the predictions with and without a signal contribution.
A clear excess of signal above the background prediction is seen at large $N_J$.
The integrated luminosity of the data sample in this Figure 
is $1~{\rm fb}^{-1}$, and a jet threshold of $50$ GeV is used. 
Square  markers show $N_J$ distributions without a requirement 
on missing energy. The effect of a missing energy requirement 
is discussed in section~\ref{met}.

\section{Robustness}

\label{robustness}

The main goal of our method is robustness against imperfections 
of the SM background modeling and detector simulation. 
By design, uncertainties in the background cross-section 
are accounted by normalizing to the yield in the forward region.
In addition, any systematic effect present in 
data should be taken into account by the background estimate, 
as long as the biases in $R_{N_J}$ ratios associated with the effect 
are a linear or slowly varying function of $N_J$.

\begin{figure}[h]
\begin{center}
\epsfig{file=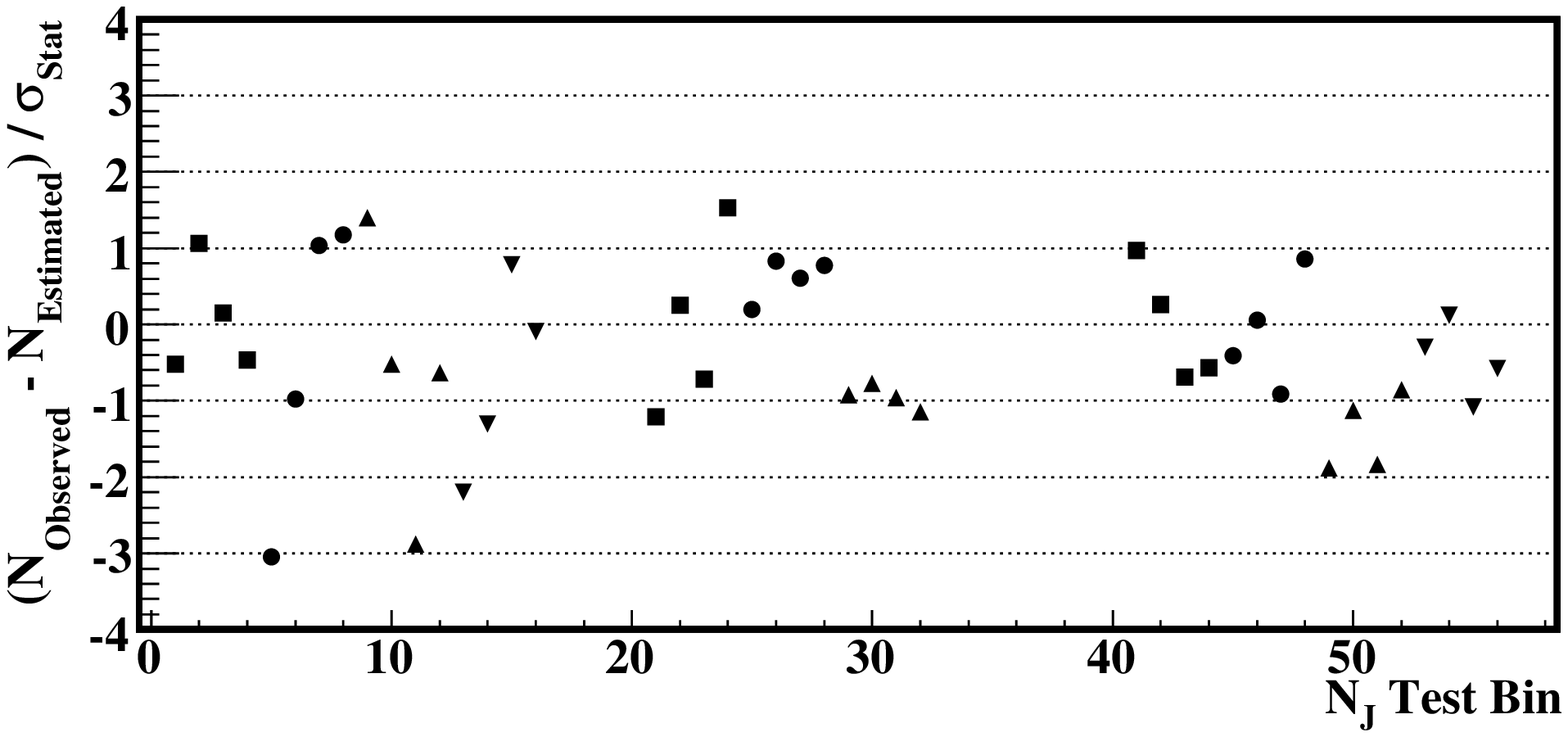,width=3.5in}
\epsfig{file=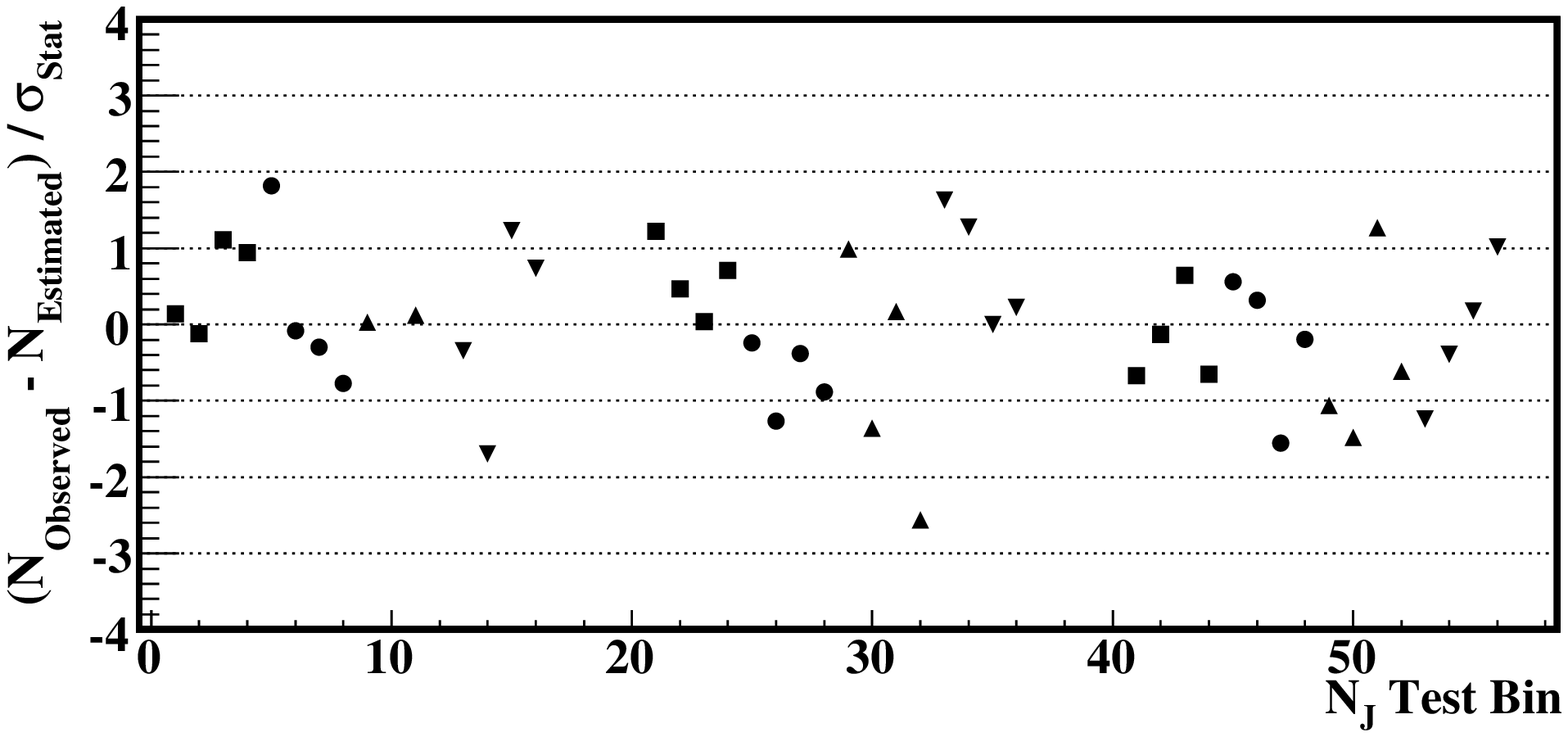,width=3.5in}
\end{center}
\caption{ Pulls between observed and estimated numbers of events
          for \zjets~(squares), \wjets~(circles),
          \gammajets~(triangle-up) and pure multi-jets~(triangle-down)
          from robustness tests in section~\ref{robustness}~(top) and 
          from tests with a requirement on \met~in section~\ref{met}~(bottom). 
          Top: $N_J$ test bins in ranges $[0;19]$, $[20;39]$ and $[40;59]$  correspond
          to tests without a requirement on \met~ consisting in changing the composition 
          of the ALPGEN sample  (\{0,2,4\} and \{1,3,5\} partons), 
          lepton/photon efficiencies 
          (over the entire $\eta$ range and in the forward region) and jet efficiencies 
          (over the entire $\eta$ range and in the forward region),  respectively. 
	  Bottom: $N_J$ test bins
          in ranges $[0;19]$, $[20;39]$ and $[40;59]$  are
          from tests with a \met~or $M_T$ requirement, 
          for different composition of the ALPGEN sample 
          (\{0,2,4\} and \{1,3,5\} partons), hypothetical holes (over the entire $\eta$ range 
	  and in the forward region) and fluctuations in jet energies (over the entire $\eta$ range 
	  and in the forward region), respectively.
          In each test pulls in the two highest $N_J$ bins are plotted. 
          (Note, pulls in these tests are correlated as tests are 
          made using events drawn from the same mock data samples.) }
\label{figure-6}
\end{figure}

To examine the robustness of our method, we present a few illustrative tests. 
In each test, a change to the mock data samples is 
made and the analysis procedure is repeated. 
The results are presented in the form of pull distributions 
in Figure~\ref{figure-6}, where only statistical uncertainties 
are used to normalize the differences between observed and 
estimated numbers of events.

The composition of the SM \zjets~sample, or other samples with a large 
number of jets, could differ from the ALPGEN predictions.
To test the effect of such mis-modeling, 
we separate the \zjets~sample into two subsamples with an even~\{0,2,4\} 
and odd~\{1,3,5\} number of ALPGEN partons and 
apply the analysis procedure to these subsamples. 
This is a particularly stringent test as it introduces drastic 
bin-to-bin variations in the $N_J$ distributions.
However, we find that the background is estimated accurately in most 
bins~[Figure~\ref{figure-6}~(top, bin range from 0 to 19)].
There are two bins, in \wjets~ and \gammajets,~where the 
observed and estimated yields differ by about 3 standard deviations. 
These biases are attributed to changes in $R_{N_J}$ associated
with the migration of events from higher to lower $N_J$ bins.
An event with $n$ jets reconstructed in the $(n - 1)$ $N_J$ bin 
has a higher probability to be a forward event, as forward 
jets are lost more often and the tag rapidity is correlated, 
although weakly, with the rapidity of the jet system recoiling against the tag.

Efficiencies for forward and central leptons are different. 
One might account for these differences by applying efficiency corrections measured from data,
but these corrections will have significant uncertainties in early data taking. 
To test the robustness of the method against mis-modeling of lepton reconstruction efficiencies, 
we change forward or central efficiencies by 30\%. 
We find that the background estimate remains 
accurate~[Figure~\ref{figure-6}~(top, bin range from 20 to 39)]~\cite{lepton_effs}.

Similarly, lepton fakes introduce background in the \zjets~and 
\wjets~channels, and photon fakes in the \gammajets~channel. 
Because the lepton and photon fake rates are expected to be a slowly varying function 
of $N_J$, background from such fakes should be accounted for accurately in our method.
When we add a small fraction of multi-jet events to the mock data samples, 
they do not significantly bias the prediction.

Significant uncertainties in the jet reconstruction efficiencies are
expected during early data taking. 
To test the robustness of the method against such inefficiencies,
jets are removed randomly with 30\% probability. 
We find that the background estimate remains accurate~[Figure~\ref{figure-6}~(top, bin range from 40 to 59)]. 
More demanding tests related to jet reconstruction efficiency and 
jet energy mis-measurements are presented below in section~\ref{met}.

We have confirmed that effects associated with 
uncertainties in the parton distribution functions are accommodated 
by our method and do not bias the background prediction.
The algorithm was also found to be robust in other tests not discussed here.

\section{Performance with \met}

\label{met}

In the results presented above, no requirement is made on missing  transverse 
energy, \met. Requiring large \met~could significantly suppress SM backgrounds, 
and it is expected to be efficient in a large class of new physics models, 
$e.g.$, $R$-parity conserving SUSY searches~\cite{susy,msugra}.
It is challenging to rely solely on \met~in analyses of early data,
because \met~is particularly difficult to model.
However, it could be useful as an additional discriminator against SM 
backgrounds in the context of our algorithm.

Unphysical sources of \met~include those associated with jet 
energy fluctuations, noise and inefficient regions of the calorimeters,
which could all be larger in the forward region. 
Our method is expected to work well with a \met~requirement, nonetheless. 
The rapidity of the tag is only weakly correlated with the rapidity of 
the jet system recoiling against the tag due to the boost 
along the beam line in the laboratory frame. 
As a result, the \met~ in the tag recoil system tends to be averaged 
over the entire rapidity coverage. 
Remaining effects can be accounted by low $N_J$ bin fits to $R_{N_J}$.

We have made a set of robustness tests with a requirement on \met~by
introducing mis-measurements and evaluating the consistency of the method's 
predictions. We require \met$> 50$~GeV~\cite{low-met-cut} for \zjets, 
\gammajets~and multi-jets.
In \wjets, the undetected neutrino is a source of genuine \met, and requiring 
\met$>50$~GeV would have little effect. Instead, we impose a requirement on
the transverse mass, $M_T$, which is constructed from \met~and the lepton's 
transverse momentum. 
Requiring $M_T > M_W + x$~GeV, where $M_W$ is the $W$ mass,
is approximately equivalent in suppressing SM
\wjets~ to requiring \met$> x$~GeV for SM \zjets.
For robustness tests in the \wjets~sample, we require $M_T > M_W + 50$~GeV.
In all four channels, the angle between the highest \pt~ jet and the missing 
transverse momentum is required to be larger than 0.15.

We repeat tests related to the ALPGEN composition of the mock 
data samples with a requirement on \met.  
To emulate the effect of holes in the detector coverage, 
we completely remove jets that fall within a cone of 
$\Delta R \equiv \sqrt{ \Delta \eta^2 + \Delta \phi^2 } < 0.7$ 
around three points in the detector, at $\eta=0$ and $\eta=\pm2$, each at $\phi=0$.
The energy of each jet is varied according to the hypothetical 
probability density function shown in Figure~\ref{figure-2} which
includes wide non-Gaussian tails.
Pulls between the observed and estimated numbers of events in high $N_J$ bins 
from these tests are shown in Figure~\ref{figure-6}~(bottom). 
Good consistency between
estimated and observed yields is seen. In these tests, the predictions are made 
based on only two $N_J$ bins: $2 \le N_J \le 3$ for \zjets~and \wjets, and $3 \le N_J \le 4$ 
for \gammajets~and multi-jets. We find that $R_{N_J}$ values in $N_J = 1$ for \zjets~and \wjets,
and $N_J = 2$ for \gammajets~and multi-jets tend to decrease after an additional
requirement on missing energy for the reason already discussed in section~\ref{robustness}. 
These bins are excluded from the background prediction procedure.
Events reconstructed in higher $N_J$ bins are less sensitive to this 
effect since the correlation between \met~ and tag rapidities is weaker in 
events with multiple jets.

The effect of a  \met~$>$~50~GeV requirement
on a search in the \zjets~sample with the jet energy mis-modeling over
the entire rapidity coverage is shown in Figure~\ref{figure-5}
in round markers. The \met~requirement suppresses the SM \zjets~rate, 
but the suppression is a function of $N_J$.
Nonetheless, our method continues to predict the background 
accurately,  and a signal excess is clearly apparent 
above the background prediction.

\section{SM \ttbarjets}

\label{ttbar}

\begin{figure*}[htbp]
\begin{center}
\begin{minipage}{7.1in}
\epsfig{file=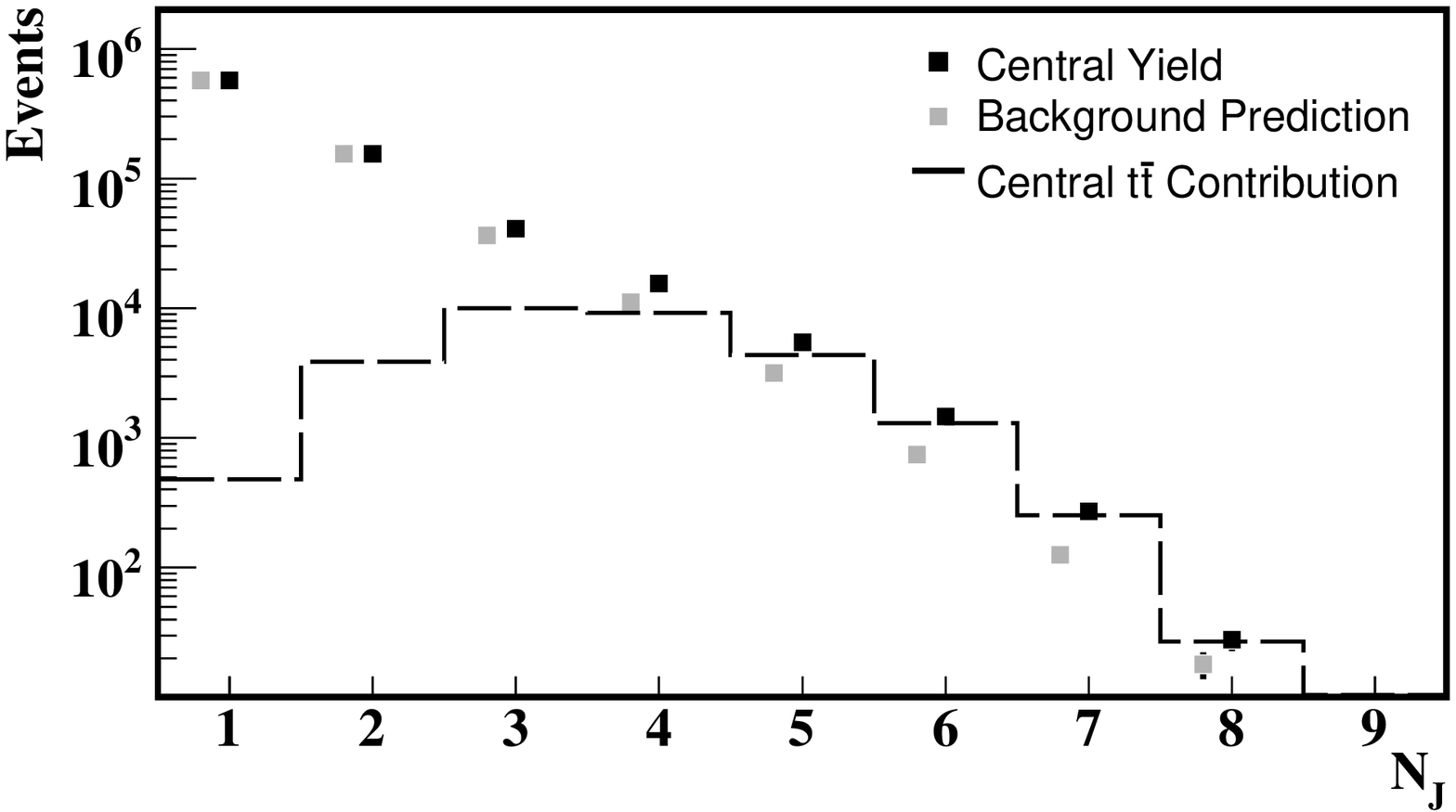,width=3.5in}
\epsfig{file=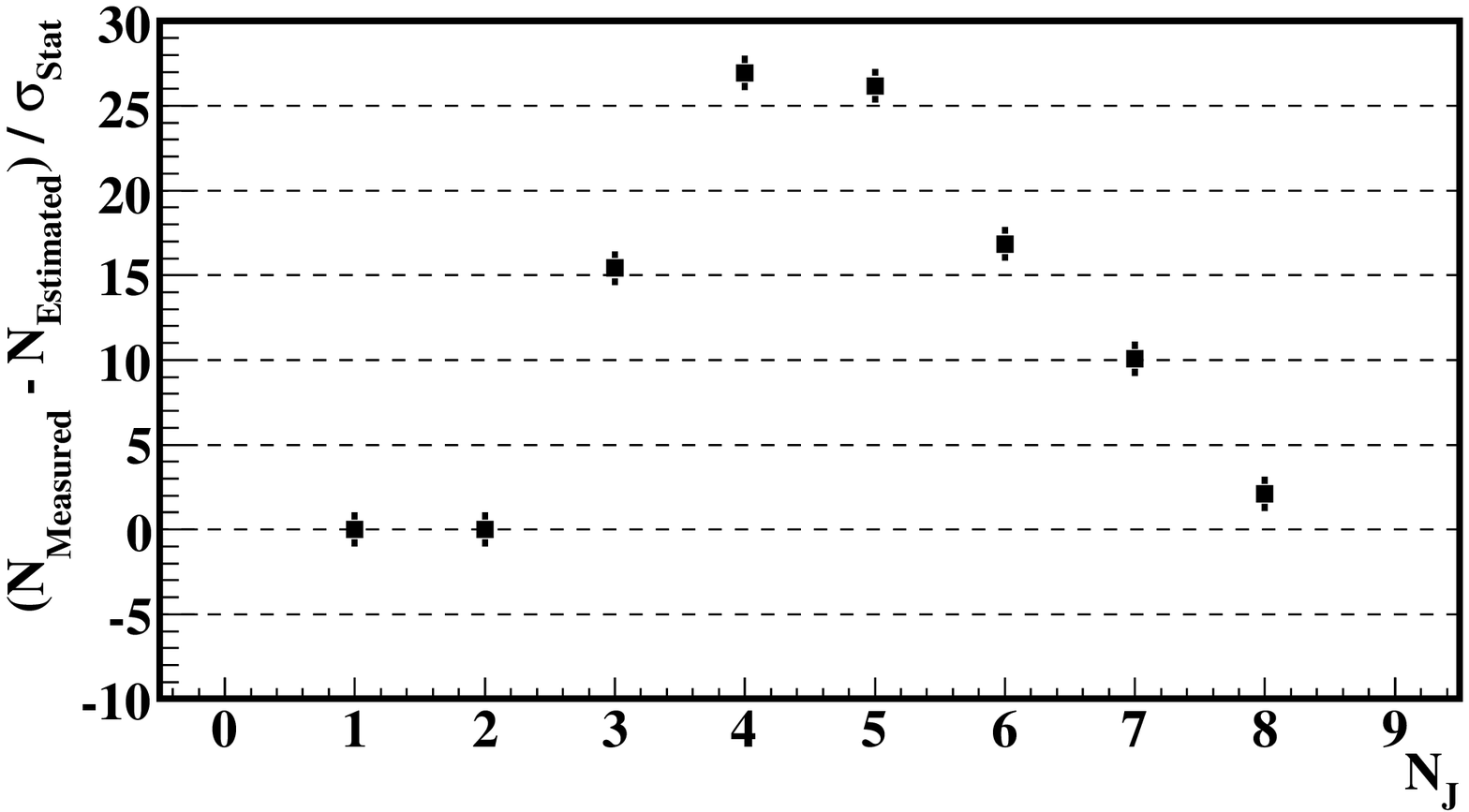,width=3.5in}
\end{minipage}
\end{center}
\caption{Results of the analysis procedure applied to the combined \wjets~and 
         \ttbarjets~sample for selection criteria defined in section~\ref{experiment}. 
	 Left: $N_J$ distributions for the combined \wjets~and \ttbarjets~sample,
         Right: pull distributions for the plots in the left column.}
\label{figure-7}
\end{figure*}

A search in the \wjets~sample is complicated by the top quark.
The \ttbarjets~process, with one of the top quarks decaying 
semileptonically and the other hadronically, produces the same signature 
as that of \wjets. Due to the large top quark mass,
the $W$ bosons from top decays tend to be produced at small rapidities, 
and they increase $R_{N_J}$ ratios over that of \wjets.

Figure~\ref{figure-7} shows results of the analysis procedure 
applied to a sample of \wjets~and \ttbarjets~events, 
where the fit to the $R_{N_J}$ distribution is made in $1 \le N_J \le 2$. 
The central yield is higher than the background prediction because of the top contribution;
the pull distribution in the right column shows the significance of the $t \bar {t}$ excess.
This demonstrates that the method works in revealing decays of massive 
particles, and it could be used to measure the \ttbarjets~cross-section.
However, \ttbarjets~complicates the search for other massive particles.

One approach to searching beyond $t \bar{t}$ would be to subtract the
\ttbarjets~ contribution, either using a prediction for its cross-section,
or an independent measurement. Another approach is to include the \ttbarjets~
background in the fit. At high $N_J$, shifts in $R_{N_J}$  
caused by \ttbarjets~ are a slowly varying function of $N_J$, so that
the method should accommodate the combined  \wjets~and \ttbarjets~
contribution in the background prediction.

\begin{figure}
\begin{center}
\vspace{-3mm}
\epsfig{file=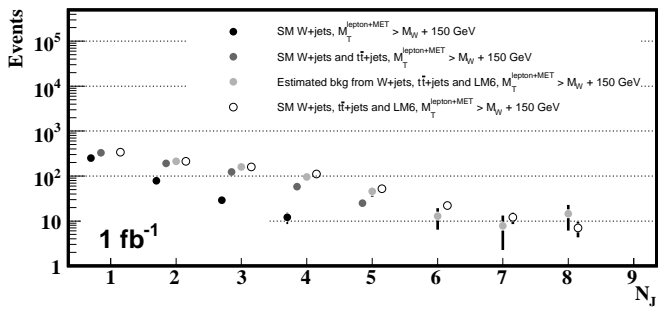,width=3.5in}
\end{center}
\caption{$N_J$~ distributions for \wjets~(black markers),
\wjets~ and \ttbarjets~(shaded markers) and a mixture of \wjets, \ttbarjets~ 
and events for LM6 mSUGRA benchmark~(open markers). Selection criteria on 
$M_T$ are given in the legend.} 
\label{figure-8}
\end{figure}

Low mass mSUGRA models are challenging for searches in $N_J$ as they produce  
$N_J$ distributions peaking in the region where the \ttbarjets~contribution 
is maximal. Figure~\ref{figure-8} illustrates this by comparing 
the central yield and prediction with and without a signal contribution.
The LM6 mSUGRA benchmark is used and the comparison is made for a sample 
size  corresponding to $1~{\rm fb}^{-1}$. A jet threshold of $50$ GeV is used, 
and a transverse mass requirement of $M_T > M_W + 150$~GeV is applied to suppress 
SM backgrounds. There is a large signal contribution at $N_J \ge 4$,
but it is not easily discernible above the central prediction 
made using $2 \le N_J \le 3$. The prediction is biased due to the residual 
\ttbarjets~contribution bridging between the \wjets~dominated low $N_J$ region 
and the signal dominated high $N_J$ region.
The $t \bar{t}$~and signal contributions together are large enough to bias 
the prediction. We discuss an alternative approach in the next section.

\section{ Search for new physics in $R_{N_J}$ }

\label{r-ratios}

In the preceding discussion, we used fits to $R_{N_J}$ to obtain a background prediction 
for the high $N_J$ distribution in central events and searched for excess signal there.
Alternatively, we can search for new physics solely in the $R_{N_J}$ distributions.
The $R_{N_J}$ ratios for heavy new particles are larger than that for SM processes,
and a search for enhancements in the high $N_J$ bins could reveal new phenomena
or provide generic bounds on it.

\begin{figure}
\begin{center}
\begin{minipage}{3.5in}
\epsfig{file=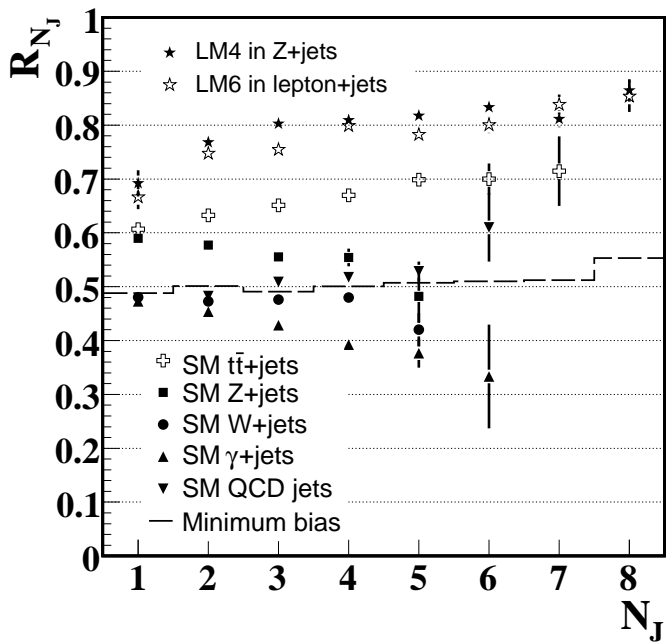,width=3.5in}
\end{minipage}
\end{center}
\caption{ 
$R_{N_J}$ distributions for minimum bias events~(track based, 
see the text), \ttbarjets~(crosses), \zjets~(squares), 
\wjets~(circles), \gammajets~(triangles-up), QCD jets~(triangles-down) 
and new physics signals~(stars). } 
\label{figure-9}
\end{figure}

Figure~\ref{figure-9} shows the $R_{N_J}$ distributions for a number of LHC processes. 
A distribution for minimum bias, $i.e.$, low \pt~scattering, events
is shown for illustration purposes, where instead of jets, tracks 
with \pt~above 3~GeV are used with the highest \pt~track providing the rapidity tag. 
Distributions for SM processes studied in this paper, \zjets, \wjets, \gammajets~ and QCD jets, 
appear approximately in the middle of the available $R_{N_J}$ range
not far from that of the minimum bias events. 
The \ttbarjets~ process contributes at higher $R_{N_J}$, due to the large top quark mass.
Distributions for LM4 and LM6 mSUGRA benchmarks in
the \zjets~and lepton+jets+\met~ channels appear at higher 
$R_{N_J}$ of about 0.8. 

\begin{figure}
\begin{center}
\begin{minipage}{3.4in}
\vspace{-3mm}
\epsfig{file=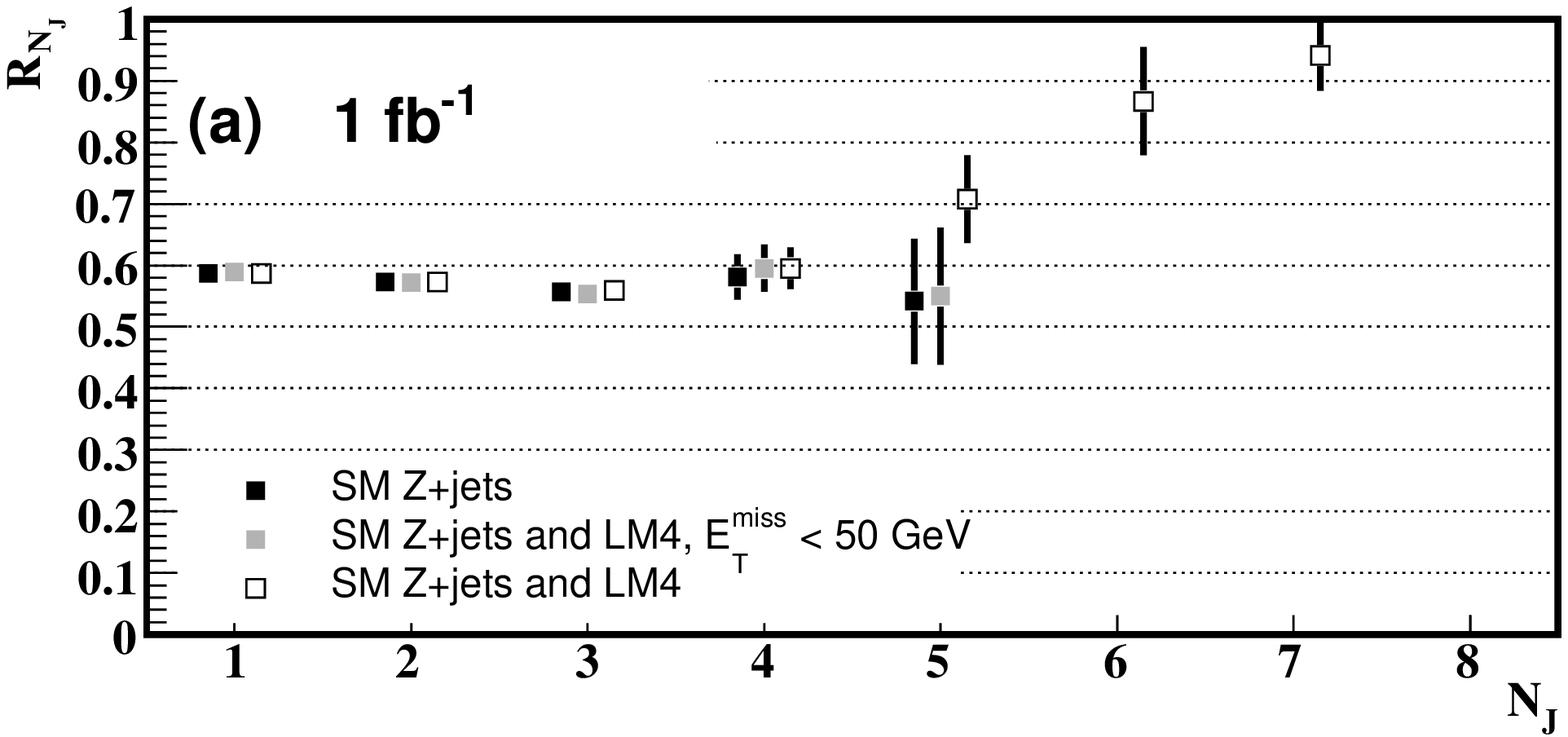,width=3.4in}
\epsfig{file=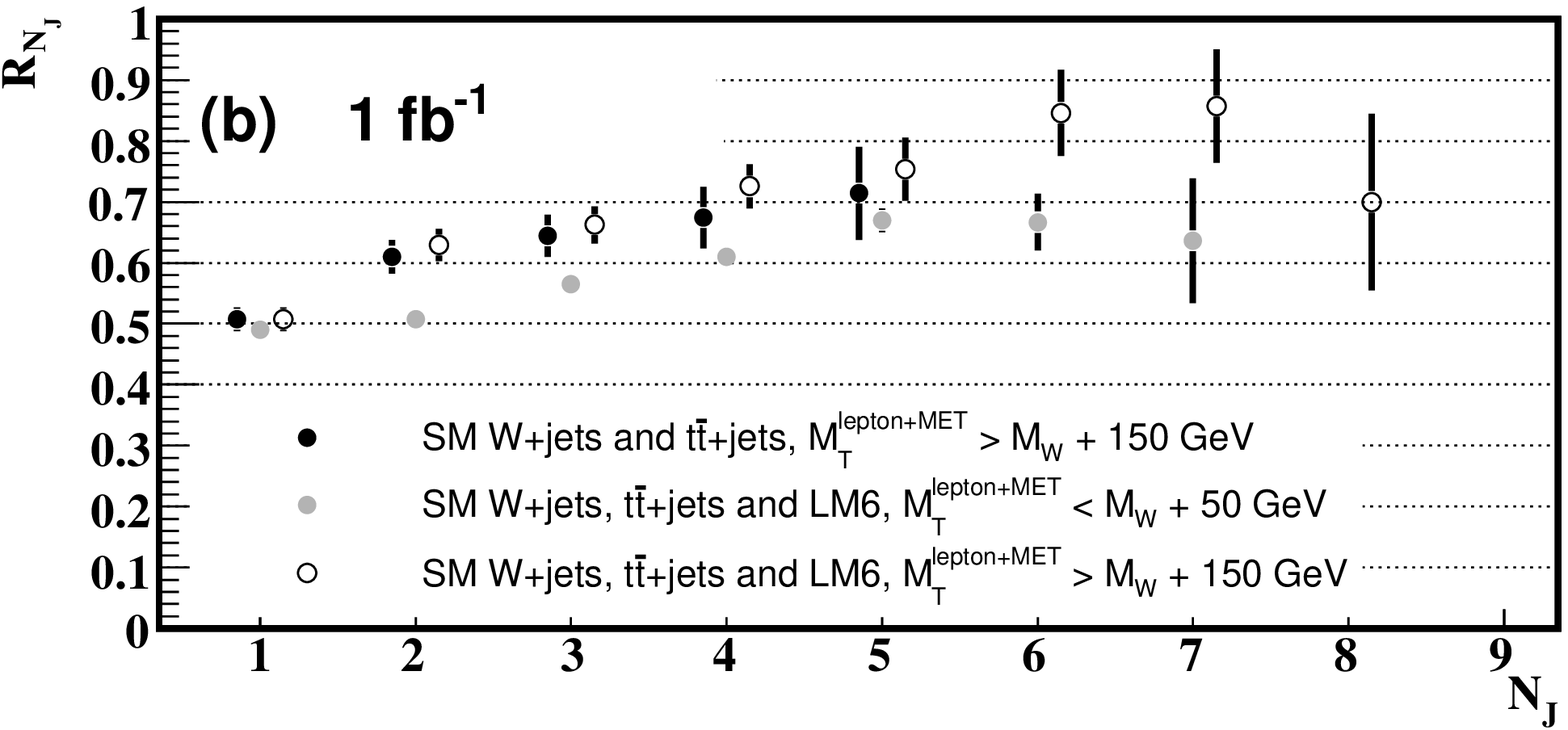,width=3.4in}
\end{minipage}
\end{center}
\caption{ 
Plot~(a): $R_{N_J}$ distributions for SM \zjets~(black markers) 
and a mixture of \zjets~and events for LM4 mSUGRA benchmark
(shaded markers: estimated central SM background, open markers: all central events).
Plot~(b):
$R_{N_J}$ distributions for \wjets~(black markers),
\wjets~ and \ttbarjets~(shaded markers) and a mixture of \wjets, \ttbarjets~ 
and events for LM6 mSUGRA benchmark~(open markers).
In both plots, a jet threshold of $50$ GeV is used; 
selection criteria on \met~ or $M_T$ are given in the legend. 
} 
\label{figure-10}
\end{figure}

The \zjets~channel has little background,
so identification of a new physics signal within it could be unambiguous.
This is illustrated in Figure~\ref{figure-10}~(a), 
where the $R_{N_J}$ distributions for SM \zjets,
with and without a new physics contribution~(LM4 mSUGRA benchmark), 
are presented. The same threshold on jet \pt~ of 50~GeV as in 
Figure~\ref{figure-5} is used. 
Black markers show the SM \zjets~ $R_{N_J}$ distribution. 
It is reproduced accurately in a sample with LM4 by 
requiring $E_T^{\rm miss} < 50$~GeV as shown in shaded markers.
Alternatively, the SM \zjets~ $R_{N_J}$ shape in the sample 
with LM4 can be obtained based on $1 \le N_J \le 3$, 
where the relative contribution from LM4 is negligible.
The new physics signal stands out clearly at $N_J \ge 5$ 
without any requirements on \met.

The \wjets~channel is complicated by the 
\ttbarjets~contribution, as discussed in section~\ref{ttbar}.
Figure~\ref{figure-10}~(b) shows the $R_{N_J}$ distribution for 
a combined \wjets~and \ttbarjets~sample, without~(black) and 
with~(shaded and open) an LM6 mSUGRA signal. 
As in Figure~\ref{figure-8}, a jet \pt~ threshold 
of 50~GeV is used and $M_T$ is required to be 
greater than $M_W + 150$~GeV to suppress SM backgrounds.
The integrated luminosity of the data sample is $1~{\rm fb}^{-1}$.
Similarly to the search in \zjets, the SM reach in $R_{N_J}$ at high 
$N_J$ can be constrained by using the sample with LM6 and 
requiring $M_T < 50$~GeV as shown in shaded markers. 
There is a large signal excess at $N_J \ge 4$, but the 
discriminating power of the search in $R_{N_J}$ in 
the lepton+jets+\met~signature for low mass mSUGRA 
models is limited by the residual \ttbarjets~contribution. 
The identification of new physics in $R_{N_J}$ producing 
larger number of jets compared to low mass mSUGRA models 
could be possible.

The search in $R_{N_J}$ is based on the distribution of tags 
in (pseudo-)rapidity in events from the same $N_J$ bin.
One can include additional information in the search from event yields 
in neighboring bins. At sufficiently 
high $N_J$ additional jets are produced via higher order 
QCD processes so that the $N_J$ distributions fall steeply in that regime.  
Selection criteria imposed on object \pt~thresholds and \met~ 
can significantly modify the $N_J$ spectra. However, 
a very general expectation is that the SM $N_J$ yields fall 
approximately exponentially at high $N_J$, while new physics can modify it.
We can use that expectation without relying heavily 
on the shape of the $N_J$ spectrum.

\begin{figure}
\begin{center}
\begin{minipage}{3.4in}
\vspace{-3mm}
\epsfig{file=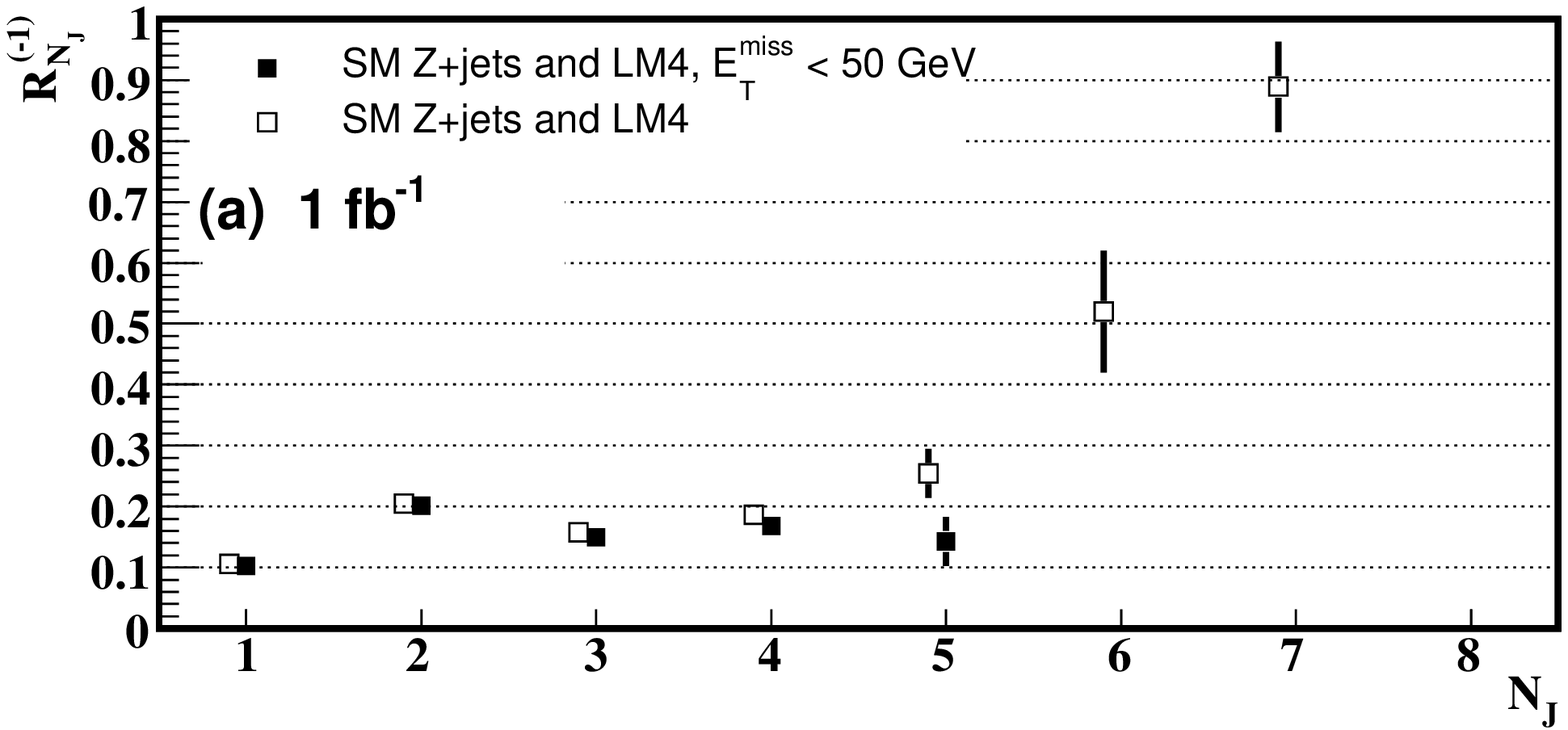,width=3.4in}
\epsfig{file=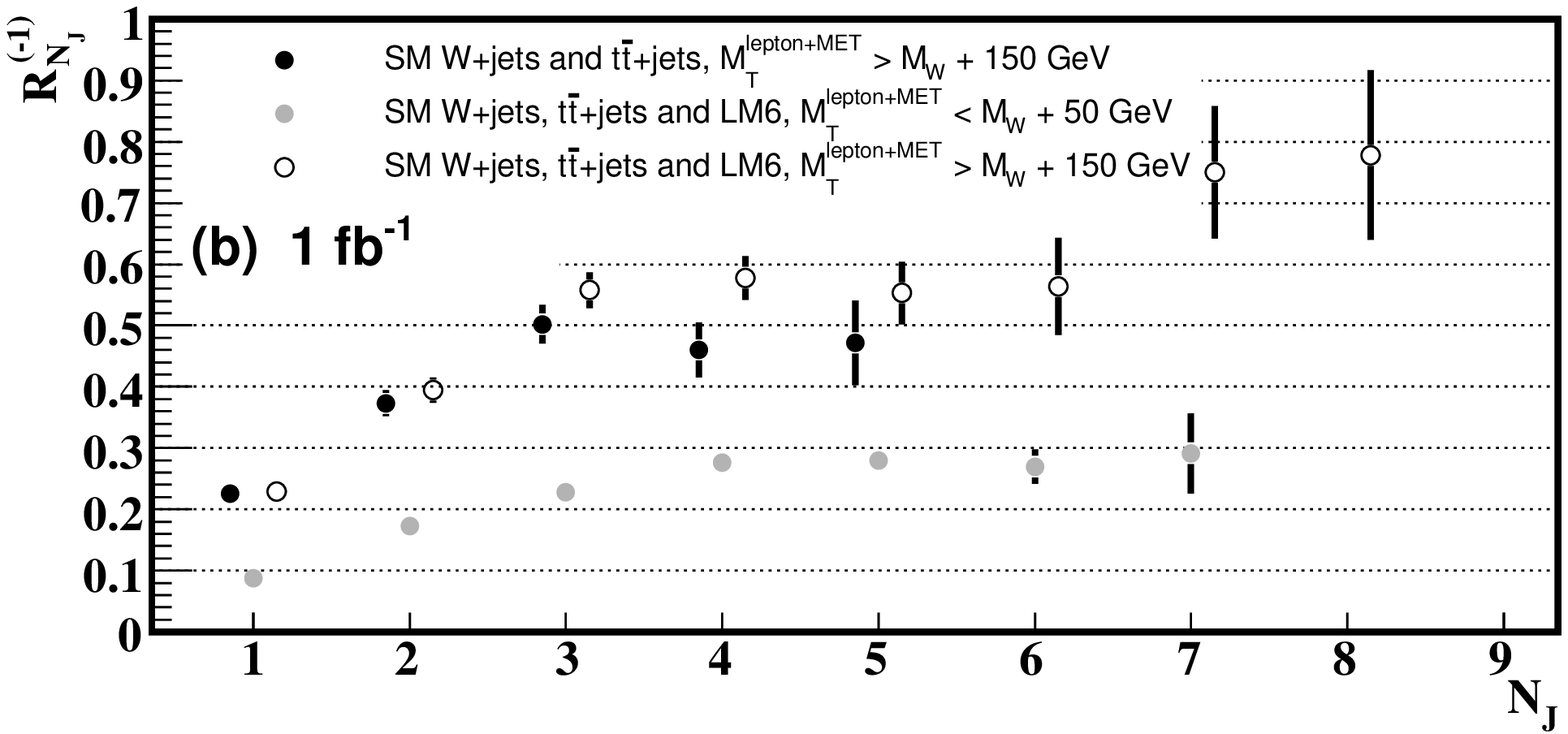,width=3.4in}
\end{minipage}
\end{center}
\caption{ 
Plot (a): $R^{(-1)}_{N_J}$ distributions for \zjets~(black markers) 
and a mixture of \zjets~and events for LM4 mSUGRA benchmark
(open markers).
Plot (b): $R^{(-1)}_{N_J}$ distributions for \wjets~and a mixture of \wjets, \ttbarjets~ 
and events for LM6 mSUGRA benchmark. 
In both plots, a jet threshold of $50$ GeV is used; 
selection criteria on \met~ or $M_T$ are given in the legend.
} 
\label{figure-11}
\end{figure}

\begin{figure}
\begin{center}
\begin{minipage}{3.4in}
\vspace{-3mm}
\epsfig{file=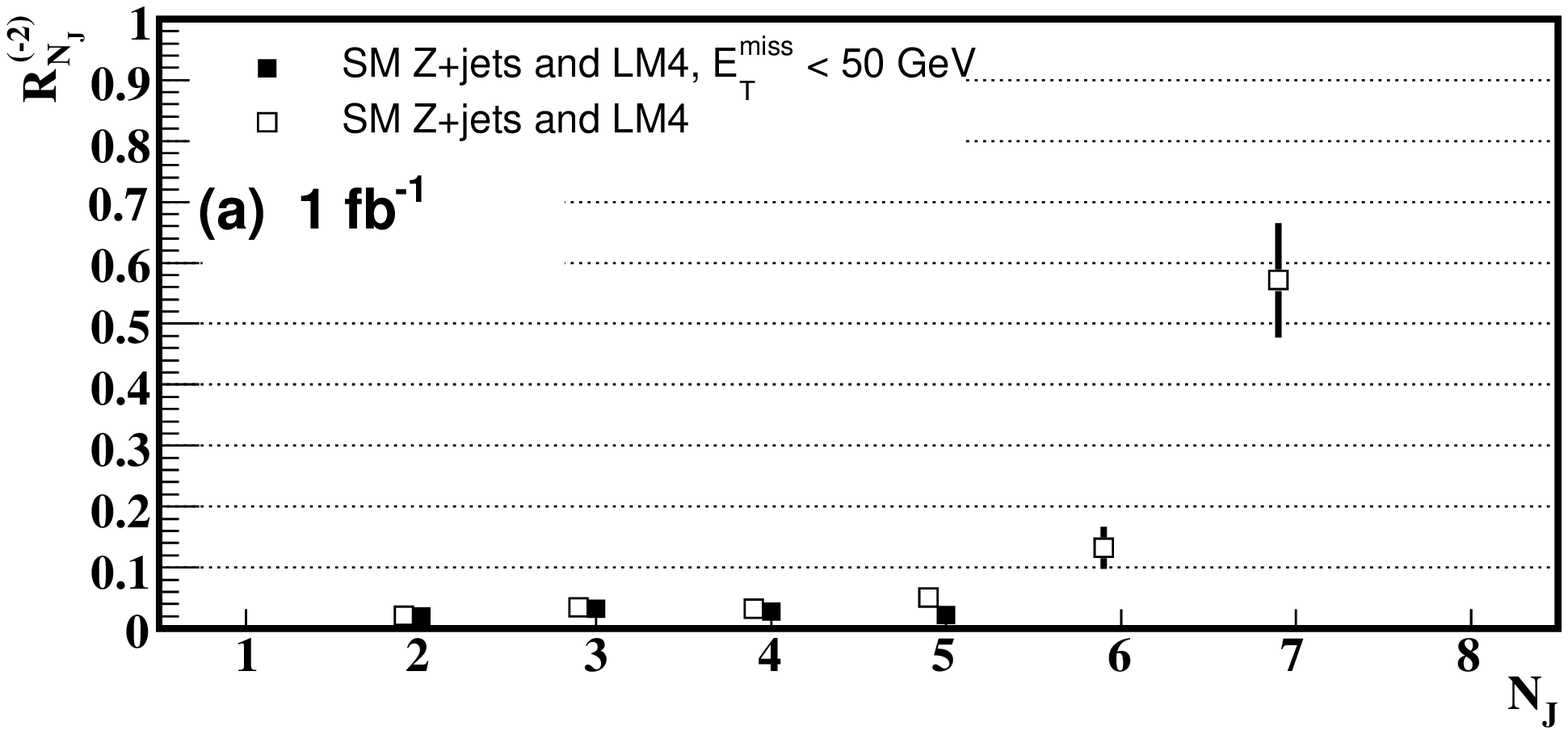,width=3.4in} 
\epsfig{file=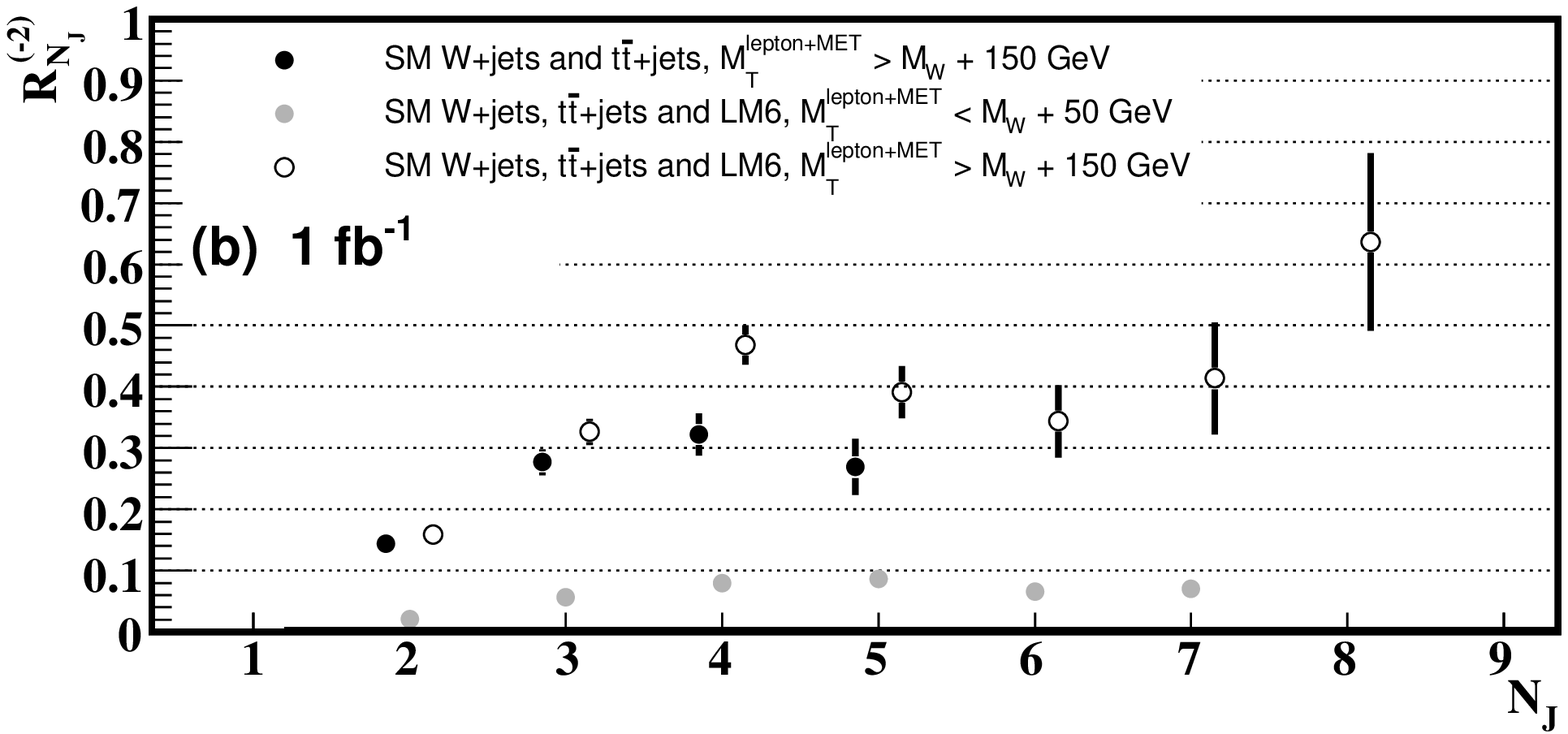,width=3.4in}
\end{minipage}
\end{center}
\caption{ 
Plot (a): $R^{(-2)}_{N_J}$ distributions for \zjets~(black markers) 
and a mixture of \zjets~and events for LM4 mSUGRA benchmark
(open markers).
Plot (b): $R^{(-2)}_{N_J}$ distributions for \wjets~and a mixture of \wjets, \ttbarjets~ 
and events for LM6 mSUGRA benchmark. 
In both plots, a jet threshold of $50$ GeV is used; 
selection criteria on \met~ or $M_T$ are given in the legend.
} 
\label{figure-12}
\end{figure}

To that end, we consider another observable 
$R^{(-1)}_{N_J} \equiv Y^{\rm Central}_{N_J}/(Y^{\rm Forward}_{N_J-1}+Y^{\rm Central}_{N_J})$,
where $Y_{N_J}$ is the event yield in the $N_J$ bin. 
It is identical to $R_{N_J}$ but in the denominator the forward yield 
in the $N_J - 1$ bin is used.
Similarly, one can define $R^{(-2)}_{N_J}$, where 
the denominator includes the forward yield in the $N_J - 2$ bin.
Figures~\ref{figure-11} and~\ref{figure-12} show $R^{(-1)}_{N_J}$ 
and $R^{(-2)}_{N_J}$ for 
the \zjets~and \wjets~samples using the previously described selection. 
The signal excess is clear and enhanced in the \zjets~sample.
For the \wjets~sample, the signal shape also has better  
separation from the background shape than in Figure~\ref{figure-10}.
These variables are less robust than $R_{N_J}$, but 
they have higher discriminating power against the background.

Using quantities like $R_{N_J}$, $R^{(-1)}_{N_J}$ or $R^{(-2)}_{N_J}$ 
could allow direct comparison across several signatures,
those considered in this paper as well as others, such as, same-sign 
or opposite-sign di-leptons, jets and \met. 
As such, they could be used to quickly perform 
a comprehensive search for new physics across multiple signatures 
in a few simple distributions.

\section{ Systematic uncertainties }

\label{systematics}

The background estimation method discussed in this paper is not subject to 
the theoretical and experimental systematic uncertainties 
usually associated with MC simulation, 
since the background shapes and normalization are measured from data. 
Instead, systematic uncertainties come from the statistical precision for 
extrapolating event yields from large to small rapidity 
and from uncertainties in the validity of a linear extrapolation  in $R_{N_J}$.
There are several sources for an extrapolation bias.

SM processes in which jets are produced via a mechanism other than 
initial or final state radiation could bias the background prediction. 
The effect of $t \bar{t}$ discussed above is an extreme example.
Di-boson production is another, 
$e.g.$, $WZ$ with a hadronic $W$ boson decay peaks at $N_J \approx 2$
in the \zjets~channel.
The cross-sections for di-boson processes can be measured, 
but even if not, they are sufficiently small so that their 
contributions are negligible.

A linear extrapolation in $R_{N_J}$ is valid only approximately. 
Large correlations between $N_J$ and 
the rapidity dependence of the tag can lead to a bias.
For example,  for $N_J = 1$ in the \gammajets~sample, 
the \pt~of the $\gamma$ used for the rapidity tag is  
directly correlated with the \pt~of the recoiling jet.
The effect of correlations can be measured by varying 
the threshold and identification requirements for 
jets, leptons, photons and \met.
Lowering thresholds will suppress sensitivity to massive new particles 
and result in a wider $N_J$ range that is background dominated.
Such background samples could be used for systematic studies such as 
comparison of alternative, $i.e.$, non-linear parameterizations and different $N_J$ 
fit ranges. Varying the $\eta$ ranges used to define forward and central events 
would have similar utility.

The usage of different, in-situ control samples is important to optimize 
and validate the final algorithm with data, and quantify its systematic 
biases. We expect that dominant systematic uncertainties 
will be associated with statistical uncertainties in such control samples.

\section{Conclusion}

\label{summary}

We have presented a new method to predict SM backgrounds within 
the context of a search for new phenomena
in final states with multiple jets: \zjets, \wjets, \gammajets~ and  multi-jets.
The fraction of central events, measured in events with few jets, 
is used to extrapolate the backgrounds measured in the forward region into the
central region for events with many jets.
This fraction of central events is identified as a new discriminator between 
SM and heavy new particles and it could be useful in any new physics search at LHC.

The method performs well in robustness tests without and with 
a requirement on the presence of significant missing transverse energy.
We have discussed systematic uncertainties associated with the method
and procedures to estimate them.
The usage of a ratio cancels many experimental uncertainties,
and the data-driven procedure avoids theoretical uncertainties.
This analysis could be performed without recourse to MC  
in early LHC data,
when robustness against imperfections of background modeling  and 
detector simulation can be a key to the discovery of new phenomena.


\begin{thebibliography}{99}
\bibitem{rapidity} Rapidity of a particle (or a jet) is defined as 
$y = \frac{1}{2}\ln(\frac{E+p_z}{E-p_z})$, where $E$ and $p_z$
are the particle's energy and the momentum component along 
the beam line. Pseudo-rapidity is $\eta = -\ln[ \tan(\theta/2)]$,
where $\theta$ is the particle's polar angle to the beam line.
\bibitem{ttbarlater} The SM \ttbarjets~ background becomes significant 
or dominant depending on requirements on the jet \pt~and \met~ in \wjets~
(lepton+jets+\met~ signature) and pure multi-jets. Due to the large mass of the top quark 
it is discussed separately in section~\ref{ttbar}.
\bibitem{other-tags}  Other definitions of tags can be made. 
For example, in the \zjets, \wjets~ and \gammajets, the highest 
\pt~ jet could alternatively be used as a tag, which 
has an advantage due to the large jet $\eta$ coverage.
\bibitem{lepton_effs} Moreover, it is not necessary to apply any lepton efficiency
corrections in our procedure as they will be accounted for in the fit
to  $R_{N_J}$ as explained earlier. The example is given to illustrate 
the insensitivity of our method to uncertainties in the lepton reconstruction 
efficiencies.
\bibitem{low-met-cut}

The small threshold for \met~ is used to retain sufficiently high yields to illustrate
the performance of the method. In section~\ref{r-ratios}, we present results with
realistic tighter selection criteria on \met~ and jet \pt.
\end{thebibliography}
\end{document}